\documentclass[acmsmall,screen]{acmart}

\acmJournal{CSUR}

\setcopyright{acmcopyright}
\copyrightyear{2022}
\acmYear{2022}
\acmDOI{XXX/XXX.XXXX}

\usepackage[utf8]{inputenc}
\usepackage[normalem]{ulem}

\usepackage{multirow}
\usepackage{tabularx}
\usepackage{array}
\newcolumntype{M}[1]{>{\centering\arraybackslash}m{#1}}
\newcolumntype{R}[1]{>{\raggedleft\arraybackslash}m{#1}}
\newcolumntype{L}[1]{>{\raggedright\arraybackslash}m{#1}}
\newcolumntype{Y}{>{\raggedright\arraybackslash}X}
\usepackage{nicematrix}

\usepackage{graphicx}
\graphicspath{{figures/}}
\usepackage{tikz}
\usepackage{pgfplots}
\pgfplotsset{compat=1.17}
\usetikzlibrary{positioning}

\newcommand*\halfcirc[1][1ex]{%
\begin{tikzpicture}
\draw[fill] (0,0)-- (90:#1) arc (90:270:#1) -- cycle ;
\draw (0,0) circle (#1);
\end{tikzpicture}}

\newcommand*\fullcirc[1][1ex]{\tikz\fill (0,0) circle (#1);}

\newcommand*\qcirc[1][1ex]{%
\begin{tikzpicture}
\draw[fill] (0,0)-- (180:#1) arc (180:270:#1) -- cycle ;
\draw (0,0) circle (#1);
\end{tikzpicture}}

\begin{document}

\hypersetup{citecolor=.}

\title{A Survey of User Perspectives on Security and Privacy in a Home Networking Environment}

\author{Nandita~Pattnaik}
\email{np407@kent.ac.uk}
\orcid{0000-0003-1272-077X}
\author{Shujun Li}
\email{S.J.Li@kent.ac.uk}
\orcid{0000-0001-5628-7328}
\author{Jason R.C. Nurse}
\email{J.R.C.Nurse@kent.ac.uk}
\orcid{0000-0003-4118-1680}
\affiliation{%
  \department{Institute of Cyber Security for Society (iCSS) \& School of Computing}
  \institution{University of Kent}
  \city{Canterbury}
  \country{UK}
  \postcode{CT2 7NP}
  }

\begin{abstract}
The security and privacy of smart home systems, particularly from a home user's perspective, have been a very active research area in recent years. However, via a meta-review of 52 review papers covering related topics (published between 2000 and 2021), this paper shows a lack of a more recent literature review on user perspectives of smart home security and privacy since the 2010s. This identified gap motivated us to conduct a systematic literature review (SLR) covering 126 relevant research papers published from 2010 to 2021. Our SLR led to the discovery of a number of important areas where further research is needed; these include holistic methods that consider a more diverse and heterogeneous range of home devices, interactions between multiple home users, complicated data flow between multiple home devices and home users, some less-studied demographic factors, and advanced conceptual frameworks. Based on these findings, we recommended key future research directions, e.g., research for a better understanding of security and privacy aspects in different multi-device and multi-user contexts, and a more comprehensive ontology on the security and privacy of the smart home covering varying types of home devices and behaviors of different types of home users.
\end{abstract}

\setcopyright{acmlicensed}
\acmJournal{CSUR}
\acmYear{2022} \acmVolume{1} \acmNumber{1} \acmArticle{1} \acmMonth{1} \acmPrice{15.00}\acmDOI{10.1145/3558095}

\ccsdesc[500]{Security and privacy~Human and societal aspects of security and privacy}

\keywords{Security, privacy, systematic literature review, user perspectives, home, networking, IoT, smart devices}

\maketitle

\section{Introduction}
\label{sec:Intro}

Modern information and telecommunication technologies (ICT) have made today's homes more connected and digitized. With the rapid advancement of artificial intelligence (AI) technologies and their use in modern homes, the more traditional term ``home network'' is increasingly replaced by more recent one ``smart home", which often covers the use of IoT (Internet of Things) home devices with ``smart'' functionalities and relying on data exchanges with the Internet. Today's smart homes are equipped with many computing/networking and smart devices, sensors, systems, and software applications. According to a 2020 report~\cite{Statista_2022_home_devices_household_worldwide2020}, the average number of connected devices in a household in most Western countries is over 7. All home devices communicate with each other and the network/Internet following a range of different protocols, while interacting with internal and external entities including home users and other individuals~\cite{Anwar2017SmartHome}. This continuously evolving home networking environment offers a multitude of benefits and opportunities to home users, but also simultaneously presents many challenges, including varied security threats and privacy issues. 

Current research in this area has a significant focus on understanding the home users' perspectives, reflecting on their awareness~\cite{zheng2018user, wickramasinghe2019survey, freudenreich2020responding, Knutzen2021Awareness}, behavior~\cite{topa2018usability, he2018rethinking,forget2016UserEngagement, Li_2020_Priviledge_Behavior, Al-Ameen2021Behaviour}, actions~\cite{Anderson2010practicing, binns2017Privacy, dupuis2018help, nthala2018rethinking} and concerns~\cite{zimmermann2019assessing, lee2020home, Almutairi2021Concerns}. In order to understand the scope and nature of these studies and ascertain the direction of future research, it is important to collate, review and analyze the relevant research in this field. Past reviews on related topics are more focused on product-centric analyses~\cite{schiefer2015smart, ho2010SecurityChoice, wickramasinghe2019survey, sarwar2018Taxonomy}, are technical and security related~\cite{heartfield2018taxonomy, alshnta_2018_sdn,rahimi_2020_fog-based}, consider the smart home in general~\cite{varghese2018framework, edu_2020_smart, yan_2020_Survey} or are purely privacy oriented~\cite{kraemer2018researching}. Reviews that covered related user perspectives studies, are either too dated~\cite{howe2012psychology, wilson2015smart}, or have a narrower scope, e.g., privacy only~\cite{kraemer2018researching}, or focused on a particular segment of home users (older users' privacy attitude)~\cite{townsend2011privacy, pal2017smart}.

Hence, our goal in this paper is to review the current research in this field and determine the areas where further research is necessary. With this in view, we need to explain two important terms, which are used throughout this paper. Firstly, we consider the term ``home'' as a relatively broader concept, covering traditional family residences, shared student accommodations, shared flats/houses, residential care homes, and nursing homes. We use the term ``home network'' to signify a network of all computing and connected devices in a home that may or may not be considered smart devices. When we use the term ``smart home'' or ``smart home network'', we refer to a slightly narrower concept, i.e., a home network that includes at least one or more smart devices, which can be controlled from a smart device or a personal computer.\footnote{A similar definition can be found at \url{https://www.lexico.com/definition/smart_home}.} Note that smart mobile devices and wearables may not be considered as smart devices by some home users and vendors, so the term ``smart device'' and ``smart home'' can have different meanings for different people. Secondly, ``user perspectives'' in our paper will incorporate a broad range of topics including mainly the following: 
\begin{itemize}
\item[UP1:] home users' behaviors, awareness, perceptions, attitudes, practices and concerns relating to security and privacy of the home network;

\item[UP2:] the relevant contexts in which UP1 occur or change;

\item[UP3:] effects of different demographic factors such as age, gender on UP1; and

\item[UP4:] theoretical frameworks that can help explain UP1.
\end{itemize}

According to the systematic meta-review we conducted as part of the research reported in this paper, covering 52 literature reviews published between 2000 -- 2021 (Table~\ref{Tab:relatedWork}), only 10 papers cover some (mostly an incomplete set of) topics related to `\textbf{user perspectives}' and only one paper~\cite{howe2012psychology} published in \citeyear{howe2012psychology}, covers a more complete discussion of related topics. The results of our meta-review indicate a gap of more recent literature review on security and privacy of smart home systems, from a home user's perspectives.

Our work was conducted in a two-staged approach. In the first stage, we conducted the above-mentioned systematic meta-review, to have a better understanding of related literature review papers. The results of the meta-review helped shape the methodology of a subsequent systematic literature review (SLR) in the second stage.

Key contributions of our work and noteworthy findings from our SLR are summarized below.
\begin{enumerate}
\item Methodologically, we used a meta-review to systematically examine past literature reviews and to facilitate design of a follow-up SLR, which is a review method that has not been used in similar past work.

\item Compared with past literature review papers, our SLR has the most comprehensive coverage of different user perspectives related to security and privacy of smart home systems, and covers more recent research papers from 2010 until 2021.

\item A number of key findings and recommendations for future research directions are obtained from our SLR, many of which have not been discussed in previous review papers. We list these below.

    \begin{itemize}
        \item The hybrid nature of a modern home with \emph{multiple (inter-)connected devices}, including different types of traditional and ``smarter'' devices, is still under-studied.

        \item The existence of \emph{multiple and different types of home users} has not been sufficiently considered.
        
        \item Study of home network and related security and privacy issues in different context and various cross-contextual effects are still under-researched areas.
        
        \item More research is needed to better understand \emph{data flows} across multiple devices and users and in different contexts.
        
        \item Some demographic factors such as location and income of home users are less studied than others.
        
        \item Research on home users' perspectives in relation to the security and privacy of the home network is predominately focused on a small number of smart devices, e.g., smart speakers and smart cameras.
        
        \item More advanced theoretical and conceptual frameworks, such as smart home security ontologies, need to be developed to support a more holistic view of security and privacy aspects of a home network environment.
    \end{itemize}
\end{enumerate}

The rest of this paper is structured as follows. Section~\ref{sec:GenMethod} discusses the general methodology followed to conduct both studies. Sections~\ref{sec:metaReview} and \ref{sec:SLR} explain the methodology and results of the two stages (the meta-review and the SLR), respectively. Note that our meta-review plays the role of the related work section of a more traditional literature review paper. Section~\ref{sec:discussions} summarizes our main findings and recommendations for future research, while the later section concludes the paper.

\section{General Methodology}
\label{sec:GenMethod}

\begin{figure}[!htp]
\includegraphics[width=\linewidth]{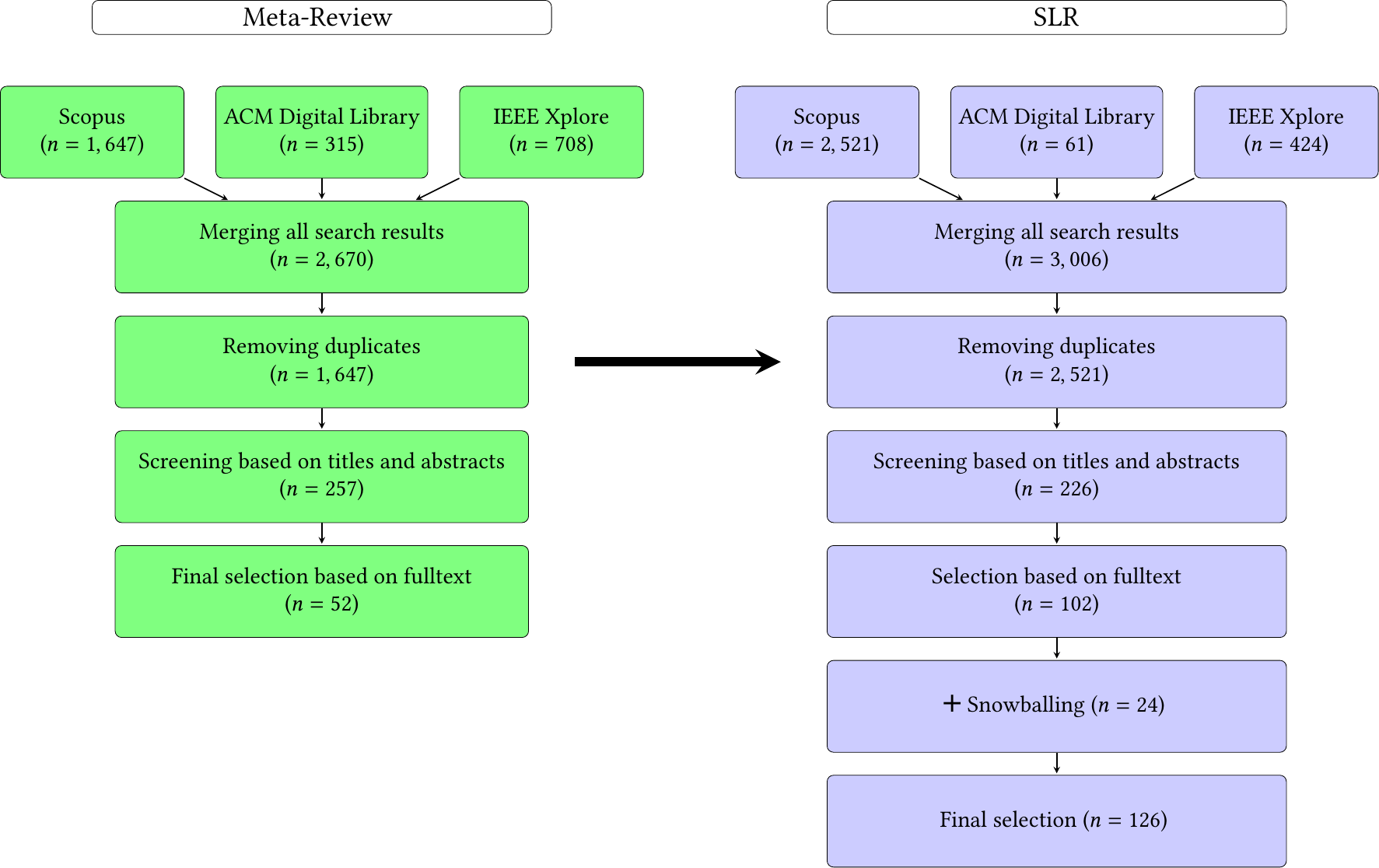}
\caption{The PRISMA procedures used for the meta-review and the SLR}
\label{fig:PRISMA}
\Description{This PRISMA flow diagram covers both stages of our work: the left hand side shows the selection process for the meta-review, depicting the selection of 52 literature review papers from a total of 2,670 papers, and the right hand side illustrates the selection process of the SLR, where 126 papers were selected from a total of 3,006 papers after removal of duplicates, a title and abstract scan, full text assessment followed by a snowballing process which generated 24 papers.}
\end{figure}

As mentioned in the previous section, the methodology we followed is two-staged: 1) a meta-review to gather the related reviews in this field more systematically and to help formulate research questions for the SLR, and 2) an SLR to collate and analyze the original research in this field. For both the meta-review and the SLR, we followed the PRISMA (Preferred Reporting Items for Systematic Reviews and Meta-Analyses) method~\cite{moher2009preferred}, a widely used procedure for conducting SLRs. The procedures followed in both stages of our study are illustrated in Fig.~\ref{fig:PRISMA}.

The PRISMA procedure to identify eligible papers for the systematic review process includes four main steps: (1) identifying pertinent records, (2) screening the selected records based on the exclusion and inclusion criteria, (3) assessing the eligible records, and (4) selecting eligible items for the final study. Since the results of the meta-review were produced systematically and contributed to the selection of our research questions for the SLR, we decided to present both the above scientific process in one flow diagram presented in Fig.~\ref{fig:PRISMA}. For ease of understanding, both these processes are color coded (blue for the SLR, and green for the meta-review) to represent two different processes. 

For both stages, a selection of three major scientific databases, Scopus, ACM Digital Library and IEEE Xplore, were considered. We decided to include Scopus as it is regarded as a very comprehensive and interdisciplinary database~\cite{Bar2018DatabaseCompare}. Note that Scopus is a product of the largest scientific publisher Elsevier, so research papers published by Elsevier are well covered by Scopus. In addition, we observed that research papers published by other mainstream publishers such as Springer and John Wiley \& Sons, Inc.\ are substantially indexed by Scopus.
We decided to include ACM Digital Library and IEEE Xplore as additional databases because ACM and IEEE are the two most important subject-specific publishers for cyber security and smart home research. Although the same query was used for all the databases, each database offered different searching tools, i.e., Scopus supported searching directly into paper title, abstract and keywords (called ``meta data''), but ACM Digital Library and IEEE Xplore did not offer such a direct search option, so we decided to search into the abstract only. The results were filtered to include only `Journal articles', `Conference proceedings', `Conference reviews', `Reviews', and `Articles in press', published in the English language. Specific keywords used, the inclusion and exclusion criteria for each of the review and the period of search, have been included in the respective methodology sections.

To store, categorize and analyze our data in both stages, we used a widely used research software system called MAXQDA (\url{http://www.MAXQDA.com/}). MAXQDA has several useful tools to support different qualitative and quantitative analysis tasks such as ``Smart coding tool'', ``Document variable analysis'', ``Visual tools'', ``Memos'', ``MaxDictio'' which were very helpful in our analysis.

\section{Meta-Review}
\label{sec:metaReview}

The meta-review aims to identify, collect, analyze and synthesize the review papers on the subject of security and privacy at home, and helps to formulate the research questions for the main SLR. The meta-review concept is similar to \emph{scope review}~\cite{munn2018systematic} that can be used as a precursor to an SLR to help guide the design of the SLR.

\subsection{Methodology}
\label{section:Methodology-metaReview}

Following our research aim, the meta-review focused on the research question: \textit{How have existing review papers in the area of security and privacy of home networking covered research on user perspectives?}

\subsubsection{Search Keywords}

Table~\ref{tab:meta-review_keywords} shows the keywords we used to conduct our search queries. The search strategy includes three main components: the first subset of keywords capture the home networking and smart home context, the second limit our searches to security and privacy-related papers, and the last one covers several typical keywords indicating the nature of the paper as a review or SoK (systematization of knowledge) paper.

\begin{table}[!htb]
\centering
\caption{The list of keywords used for the meta-review}
\label{tab:meta-review_keywords}
\begin{tabularx}{\linewidth}{c X}
\toprule
& (Home AND (Network OR Networking OR Smart OR Computer OR Computing OR Internet OR Device))\\
\midrule
AND & (Security OR Privacy)\\
\midrule
AND & (Review OR Survey OR Overview OR Systematisation OR Systematization OR Systematic OR SoK)\\
\bottomrule
\end{tabularx}
\end{table}

To follow the screening and eligibility checking steps of the PRISMA process, a set of exclusion and inclusion criteria were established. A paper meeting any one of the exclusions (or inclusion) criteria was excluded (or included). 

\subsubsection{Exclusion Criteria}
\label{subsubsec:MetaReview_Exclusion}

Papers meeting the following exclusion criteria were excluded:

\begin{itemize}
\item Papers published before 2001

\item Non-English papers

\item Papers that do not cover the home context (e.g., those covering industrial IoT)

\item Conference reviews or book chapters

\item Papers that do not cover any security or privacy factors

\item Papers which are not review papers excepting non-review papers that report any taxonomy covering security, privacy or user perspectives of the home network
\end{itemize}

\subsubsection{Inclusion Criteria}
\label{subsubsec:MetaReview_Inclusion}

After excluding papers based on the exclusion criteria, those meeting at least one of the following inclusion criteria were selected:

\begin{itemize}
\item Papers focusing on issues and solutions in the area of security and privacy of the home network

\item Papers that systematically reviewed smart home and home network related products or applications
\end{itemize}

The non-review papers reporting on ontology or taxonomy relevant to the subjects were considered as pseudo-reviews and were included in the study because they normally systematically conceptualize relevant topics.

In the following, we present the results of the meta-review.

\subsection{Results \& Discussion}
\label{MetaReviewResult}

Figure~\ref{fig:PRISMA} (colored in green) shows the results from each step of the meta-review. The initial searches gave us 2,670 papers in total. After removing duplicates, we had 1,647 papers to screen. Following the exclusion criteria in Section~\ref{subsubsec:MetaReview_Exclusion}, we ended up with 257 papers for further screening. After a scan of those papers' abstracts, introduction and conclusion sections and applying our inclusion criteria presented in Section~\ref{subsubsec:MetaReview_Inclusion}, 52 papers were included for final analysis. The screening and filtering steps gave us a clear indication of reviews that were conducted in the past on relevant topics, helping to inform the second stage of our work (i.e., the SLR). Table~\ref{Tab:relatedWork} presents an overall comparison of related study with the current research. Note that although all included papers are review papers, only some can be considered SLR (i.e., others were not done following a systematic approach). Table~\ref{tab:ReviewTheme} depicts the thematic categorization of the 52 papers in our study based on the broad overall theme that the reviews conveyed.

\begin{table}[!t]
\caption{Comparison of the current study with related work}
\label{Tab:relatedWork}
\small
\begin{NiceTabular}{Wc{0.09\linewidth}c*{14}{c}}[hvlines]

\Block{3-1}{\rotate\textbf{Year(s)}} & \Block{3-1}{\rotate\textbf{Reference(s)}}
& \Block{1-12}{\textbf{Subject Focus}} & & & & & & & & & & &
& \Block{3-1}{\rotate\textbf{SLR}} & \Block{3-1}{\rotate\textbf{Meta-Review}}\\

& &
\Block{2-1}{\rotate\textbf{Security}} & \Block{2-1}{\rotate\textbf{Privacy}} & \Block{2-1}{\rotate\textbf{Smart Home}}
& \Block{1-9}{\textbf{User Perspectives}} & & & & & & & &
& &\\

& & & & &
\rotate\textbf{Holistic} & \rotate\textbf{Multi-Device } & \rotate\textbf{Awareness} & \rotate\textbf{Concerns} & \rotate\textbf{Behavior} & \rotate\textbf{Multi-User} & \rotate\textbf{Demographics } & \rotate\textbf{Contextual} & \rotate\textbf{Theoretical}
& &\\

\Block{6-1}{2000--2016} & \cite{lopez_2014_ambient} & & $\checkmark$ & $\checkmark$ & & & & & & & & & & &\\
& \cite{kermani_2013_emerging, brezovan_review_2013} & $\checkmark$ & & $\checkmark$ & & & & & & & & & & &\\
& \cite{townsend2011privacy} & & $\checkmark$ & & $\checkmark$ & & & & & & & $\checkmark$ & & $\checkmark$ &\\
& \cite{howe2012psychology} & $\checkmark$ & $\checkmark$ & $\checkmark$ & & & $\checkmark$ & $\checkmark$ & $\checkmark$ & & $\checkmark$ & & $\checkmark$ & &\\
& \cite{wilson2015smart} & & $\checkmark$ & $\checkmark$ & & & & & & & & & & $\checkmark$ &\\
& \cite{yusif2016older} & & $\checkmark$ & $\checkmark$ & & & & $\checkmark$ & & & & $\checkmark$ & & $\checkmark$ &\\

\Block{5-1}{2017} & \cite{batalla2017secure, marksteiner_2017_overview, abrishamchi_2017_side} & $\checkmark$ & & $\checkmark$ & & & & & & & & & & &\\
& \cite{ghiglieri2017context, alam_2017_review} & $\checkmark$ & $\checkmark$ & $\checkmark$ & & & & & & & & & & &\\
& \cite{alotaibi_2017_analysis} & $\checkmark$ & $\checkmark$ & $\checkmark$ & & & $\checkmark$ & & & & & & $\checkmark$ & &\\
& \cite{pal2017smart} & $\checkmark$ & $\checkmark$ & $\checkmark$ & & & & & & & & & & $\checkmark$& \\
& \cite{barth2017privacy} & & $\checkmark$ & $\checkmark$ & & & $\checkmark$& $\checkmark$ & & & & & $\checkmark$ & $\checkmark$ &\\

\Block{4-1}{2018} & \cite{heartfield2018taxonomy} & $\checkmark$& $\checkmark$ & $\checkmark$& & & & & & & & &  & & \\
& \cite{alshnta_2018_sdn, ghazali_2018_security, gamundani_2018_overview} & $\checkmark$ & &$\checkmark$ & & & & & & & & & & &\\
& \cite{safavi_2018_cyber, varghese2018framework} & $\checkmark$ & $\checkmark$ & $\checkmark$ & & & & & & & & & & &\\
& \cite{kraemer2018researching} & & $\checkmark$ & $\checkmark$ & & & & & & & & $\checkmark$ & $\checkmark$ & $\checkmark$ &\\

\Block{3-1}{2019} & \cite{almusaylim_2019_review, kuyucu2019security} & $\checkmark$ & $\checkmark$ & $\checkmark$ & & & & & & & & $\checkmark$ & & &\\
& \cite{michler2019trust, alrawi_sok_2019} & $\checkmark$ & & $\checkmark$ & & & & & & & & & & $\checkmark$ &\\
& \cite{williams2019survey, GuptaSingh_2019_cyber, gupta_2019_survey, nadargi_2019_novel, chakraborti_2019_review} & $\checkmark$ & & $\checkmark$ & & & & & & & & & & &\\

\Block{5-1}{2020} & \cite{fatima_2020_home, davis_vulnerability_2020} & $\checkmark$ & & $\checkmark$ & & & & & & & & & & &\\
& \cite{liao_2020_security, talal_2019_smart, rahimi_2020_fog-based, sarhan_systematic_2020, chhetri2020identifying} & $\checkmark$ & & $\checkmark$ & & & & & & & & &  & $\checkmark$ &\\
& \cite{moniruzzaman_2020_blockchain, yan_2020_Survey} & $\checkmark$ & $\checkmark$ & $\checkmark$ & & & & & & & & & & &\\
& \cite{ray_2020_sensors} & $\checkmark$ & $\checkmark$ & $\checkmark$ & & & & & & & & & & $\checkmark$ &\\
& \cite{edu_2020_smart} & $\checkmark$ & $\checkmark$ & $\checkmark$ & & & & & & & & & & $\checkmark$ & \\

\Block{5-1}{2021} & \cite{defranco2021smart, bolton2021security, sun2021systematic, gochoo2021towards} & $\checkmark$ & $\checkmark$ & $\checkmark$ & & & & & $\checkmark$ & & & & & $\checkmark$ &\\
& \cite{chen2021survey} & & $\checkmark$ & $\checkmark$ & $\checkmark$ & & & & & & & & & & \\
& \cite{aljanah2021survey, mohammad2021access} & $\checkmark$ & & $\checkmark$ & & & & & & & & & & &\\
& \cite{Philip_2021_Internet} & $\checkmark$ & $\checkmark$ & $\checkmark$ & & & & & & & & & & &\\
& \cite{li2021motivations} & $\checkmark$ & $\checkmark$ & $\checkmark$ & & & $\checkmark$ & $\checkmark$ & & & & & & &\\

2022 & \textbf{Our work} & $\checkmark$ & $\checkmark$ & $\checkmark$ & $\checkmark$ & $\checkmark$ & $\checkmark$ & $\checkmark$ & $\checkmark$ & $\checkmark$ & $\checkmark$ & $\checkmark$ & $\checkmark$ & $\checkmark$ & $\checkmark$\\
\end{NiceTabular}
\end{table}

\begin{table}[!htb]
\caption{Categorization of review papers into different categories (period: 2000--2021)}
\label{tab:ReviewTheme}
\footnotesize
\begin{tabular}{|M{2cm}*{6}{|M{1.4cm}}|c|}
\hline
\textbf{Category} & \rotatebox[origin=c]{45}{\textbf{2000--2016}} & \rotatebox[origin=c]{45}{\textbf{2017}} & \rotatebox[origin=c]{45}{\textbf{2018}} & \rotatebox[origin=c]{45}{\textbf{2019}} & \rotatebox[origin=c]{45}{\textbf{2020}} & \rotatebox[origin=c]{45}{\textbf{2021}}&\textbf{(\#)}\\

\hline
Security \& privacy in general & \cite{lopez_2014_ambient} & \cite{batalla2017secure,marksteiner_2017_overview} & \cite{alshnta_2018_sdn, barriga_2018_security} & \cite{almusaylim_2019_review, kuyucu2019security, michler2019trust, williams2019survey} & \cite{fatima_2020_home, liao_2020_security, talal_2019_smart, Philip_2021_Internet, rahimi_2020_fog-based, sarhan_systematic_2020} & \cite{defranco2021smart, bolton2021security, chen2021survey, sun2021systematic}
& 19\\
\hline
User perspectives & \cite{townsend2011privacy, howe2012psychology, wilson2015smart, yusif2016older} & \cite{alotaibi_2017_analysis, barth2017privacy, pal2017smart} & \cite{kraemer2018researching} & \cite{michler2019trust} & -- & \cite{li2021motivations}&10\\
\hline

Threats and attacks & \cite{kermani_2013_emerging} &
\cite{abrishamchi_2017_side, batalla2017secure} & \cite{heartfield2018taxonomy} & \cite{alrawi_sok_2019, GuptaSingh_2019_cyber} & \cite{davis_vulnerability_2020, liao_2020_security, chhetri2020identifying} & -- & 9\\
\hline

Security \& privacy solutions & -- & -- & \cite{alshnta_2018_sdn, gamundani_2018_overview, ghazali_2018_security, safavi_2018_cyber} & \cite{ chakraborti_2019_review, kuyucu2019security, nadargi_2019_novel, gupta_2019_survey} & \cite{rahimi_2020_fog-based, moniruzzaman_2020_blockchain, sarhan_systematic_2020} &\cite{aljanah2021survey, mohammad2021access} &13\\
\hline

Smart devices & \cite{kermani_2013_emerging, brezovan_review_2013} & \cite{ghiglieri2017context, alam_2017_review} & \cite{varghese2018framework} & -- & \cite{ray_2020_sensors, edu_2020_smart, yan_2020_Survey} & \cite{gochoo2021towards} & 9\\
\hline
\end{tabular}
\end{table}

\subsubsection{Security \& Privacy in General}

Nineteen papers covered by the meta-review focus on various common issues arising from security and privacy problems in a smart home. \citeauthor{defranco2021smart}'s paper~\cite{defranco2021smart} covered the broad theme of general research avenues around smart home. While \citeauthor{kuyucu2019security}~\cite{kuyucu2019security} reviewed papers relating to both privacy and security related issues and solutions, \citeauthor{liao_2020_security}~\cite{liao_2020_security} explored security problems, challenges, techniques used, and solutions available from a mobile computing point of view. \citeauthor{talal_2019_smart}~\cite{talal_2019_smart} focused on security issues of tele-medicine environment exploring smart home issues and solutions in general and hardware sensors, protocols, wireless network, security architecture, in particular. \citeauthor{Philip_2021_Internet}~\cite{Philip_2021_Internet} published a similar study on home health monitoring systems. \citeauthor{bolton2021security}~\cite{bolton2021security} surveyed the security and privacy challenges of virtual assistants such as `Siri' by Apple. \citeauthor{barriga_2018_security}~\cite{barriga_2018_security} discussed existing security mechanisms and approaches in a smart home automation system, while \citeauthor{batalla2017secure}~\cite{batalla2017secure} analyzed the security requirements and countermeasures used in a general IoT architecture and suggested an extendable home area network with multi-level privacy and security to be managed by a trusted external actor to lessen the burden on home users. Other topics covered in some reviews include smart home data protection issues~\cite{chen2021survey} and security issues in different layers of smart home network~\cite{sun2021systematic}.

Three papers chose to focus on specific areas in the smart home security. These include, Ambient Intelligence (AmI) applications and their privacy issues by \citeauthor{lopez_2014_ambient}~\cite{lopez_2014_ambient}, major wireless protocols such as Zigbee and Z-wave by \citeauthor{marksteiner_2017_overview}~\cite{marksteiner_2017_overview}, and, and security and protection mechanism in face detection techniques by \citeauthor{fatima_2020_home}~\cite{fatima_2020_home}. Papers also looked into general smart home related topics, i.e., \citeauthor{williams2019survey}~\cite{williams2019survey} who discussed on smart home automation and trust building factors, Whereas, \citeauthor{michler2019trust}~\cite{michler2019trust} investigated research work on trust-building factors in consumer IoT products in four areas including smart home. Reviews also considered alternative approaches to different security issues in a home network such as utilizing fog-computing architecture~\cite{rahimi_2020_fog-based}, smart home safety and security using Ardunio platform~\cite{sarhan_systematic_2020} and using software defined networking (SDN)~\cite{alshnta_2018_sdn} to control home network security.

\subsubsection{User Perspectives}

We identified ten papers that reviewed the literature on privacy and security issues from home users' perspectives. Out of all 10 papers,  \citeauthor{howe2012psychology}~\cite{howe2012psychology}'s research was the only paper which touched upon many of the areas we wanted to review, although it is a very old review (\citeyear{howe2012psychology}). This research focused on the psychology and factors influencing users' security behaviors and decisions. It explored various demographic characteristics, information sources for home users, users' understanding of security risks, their perception of security behaviors and defensive security actions.

\citeauthor{wilson2015smart}~\cite{wilson2015smart} analyzed  150 relevant papers and organized their findings into three broad themes of 1) growth of smart home from `functional', `instrumental' and `socio-technical' viewpoints, 2) users and their use of smart home with the subcategories of `prospective users', `interactions and decisions' and `technology in the home', and 3) challenges in a smart home covering hardware and software, design and domestication issues. \citeauthor{alotaibi_2017_analysis}~\cite{alotaibi_2017_analysis} reviewed research work on security awareness and education amongst home users recommended an individualized approach to provide information to users based on their existing awareness level. \citeauthor{michler2019trust}~\cite{michler2019trust} analyzed user perception from the angle of trust to understand smart home adoption issues.

Two papers explored the elderly users' prospective on using smart home. \citeauthor{pal2017smart}~\cite{pal2017smart}'s systematic review on the elderly users' perspective on smart homes, observed that the elderly population have serious security and privacy concerns on the use of smart devices. In another systematic review, \citeauthor{yusif2016older}~\cite{yusif2016older} focused on assistive technology use by older people and noted that 34\% of the articles examined, recognized privacy as a major concern for the older adults.

The other four papers differed widely in their topic choices. After having systematically reviewed privacy-related papers, \citeauthor{kraemer2018researching}~\cite{kraemer2018researching} concluded that contextual privacy at home and privacy behaviors were very under-researched areas. The concept of privacy paradox and related studies were covered in two reviews~\cite{townsend2011privacy, barth2017privacy}. \citeauthor{li2021motivations}~\cite{li2021motivations} conducted a systematic review to find the motivation, barriers and risk of smart home adoption from a consumer's point of view.

\subsubsection{Threats and Attacks}

We found nine papers reviewing various aspects of security threats, attack types. \citeauthor{abrishamchi_2017_side}~\cite{abrishamchi_2017_side} discussed different types of side-channel threats in a smart home, giving a categorical view of different devices and systems layers in a smart home that leaks private data. \citeauthor{heartfield2018taxonomy}~\cite{heartfield2018taxonomy} on the other hand, produced a taxonomic view of the possible cyber-physical threats and their impact on smart home. \citeauthor{alrawi_sok_2019}~\cite{alrawi_sok_2019} produced a systematized view of the research literature on smart home security arena, under 4 major categories: `device', `mobile application', `cloud endpoint' and `communication'. Papers were further sub-categorized into `Attack vector', `Mitigation', and `Stakeholder' and evaluated with 45 IoT devices to identify the research gaps. Researchers also explored the security vulnerabilities by different IoT devices~\cite{davis_vulnerability_2020} and smart home devices~\cite{chhetri2020identifying}. \citeauthor{batalla2017secure}~\cite{batalla2017secure} classified the security threats in home area network devices by referring to the 3-layered approach of ENISA (European Union agency for Cybersecurity), i.e., perceptual, network and the application layer attack and  physical attacks. Implantable, wearables and embedded sensors were another topic of discussion~\cite{kermani_2013_emerging} along with security measures in mobile computing~\cite{liao_2020_security}. 

\subsubsection{Security \& Privacy Solutions}

Thirteen papers focused on solutions to specific security and privacy problems in a smart home. \citeauthor{chakraborti_2019_review}~\cite{chakraborti_2019_review} reviewed research work on software solutions and embedded solutions in a smart home. Two of the papers \cite{moniruzzaman_2020_blockchain, safavi_2018_cyber} reviewed blockchain-based solutions for addressing security and privacy problems of a smart home. Three papers focused on different areas of smart home solutions, including current fog-based literature~\cite{rahimi_2020_fog-based}, research on software-defined networks (SDN)~\cite{alshnta_2018_sdn} and smart home safety and security systems focusing specifically based on the Arduino platform~\cite{sarhan_systematic_2020}. \citeauthor{kuyucu2019security}~\cite{kuyucu2019security} reviewed papers both on issues and solutions. We found six papers reviewing the existing literature on authentication schemes and various security threats to them. Four papers~\cite{gamundani_2018_overview, nadargi_2019_novel, aljanah2021survey, mohammad2021access} provided a general classification of different authentication schemes, threats and attacks on IoT devices in a smart home. Other such as \citeauthor{ghazali_2018_security}~\cite{ghazali_2018_security} extensively reviewed the biometric factors reflecting on the authentication mechanisms inside a smart home whereas, \citeauthor{gupta_2019_survey}~\cite{gupta_2019_survey} focused on a comparative study of different encryption algorithms used in IoT platform and proposed solutions to increase energy efficiency of various systems.

\subsubsection{Smart Home Devices}

Nine papers covered in the meta-review focused on smart home devices, without looking at the system or user level aspects where multiple devices form a home network. \citeauthor{varghese2018framework}~\cite{varghese2018framework} reviewed research on security and privacy of different categories of smart home devices. \citeauthor{ray_2020_sensors}~\cite{ray_2020_sensors} provided an in-depth discussion on IoT-based biosensors. Four other papers covered other specific areas including smart TV~\cite{alam_2017_review}, hybrid broadcasting broadband TV (HbbTV) techniques~\cite{ghiglieri2017context}, video surveillance methods~\cite{brezovan_review_2013}, implantable and wearable medical devices, detectors and control systems such as temperature sensors or smoke detectors~\cite{kermani_2013_emerging}. \citeauthor{edu_2020_smart}~\cite{edu_2020_smart} reviewed research work on smart home personal assistants (SPA) to examine the main security and privacy issues, features that characterize known attacks, limitations of countermeasures. Features and challenges of smart gateway systems~\cite{yan_2020_Survey} and assisted smart home technologies for elderly people~\cite{gochoo2021towards} are two other areas of discussion under this category.

\subsubsection{Discussions \& Research Questions Identified for the SLR}

The meta-review led to several key findings.
\begin{itemize}
\item First, although ten papers covered user perspectives in different ways, the discussions have not considered papers from both the security and privacy angle. The reviews have reflected on the privacy issues~\cite{townsend2011privacy, kraemer2018researching}, educational awareness~\cite{alotaibi_2017_analysis} and general challenges users face with very little focus on security and privacy~\cite{wilson2015smart}. The ``User Perspective'' as discussed in Section~\ref{sec:Intro}, comprises a much broader domain. Only \citeauthor{howe2012psychology}~\cite{howe2012psychology} discussed a more comprehensive details of user perspectives including behaviors and practices of users, but the study is relatively old (published in \citeyear{howe2012psychology}). There is therefore a need of collating and synthesizing more studies in this growing area.

\item Second, 42 review papers collectively cover different aspects of smart home research, including specific (types of) smart home devices, privacy and security issues, and technical solutions. We noticed the absence of wider discussions and a more \textit{holistic} view of a smart home where \textit{multiple} and \textit{heterogeneous} devices are interacting with each other and home users. 

\item Third, as reported in~\cite{kraemer2018researching}, contextual aspects of security and privacy of home network have been much less studied.

\item Fourth, among the ten review papers covering user perspectives, demographic factors are mentioned in only one early review conducted by \citeauthor{howe2012psychology}~(\citeyear{howe2012psychology})~\cite{howe2012psychology}.

\item Fifth, although not a systematic review, \citeauthor{howe2012psychology}~\cite{howe2012psychology} conducted a very comprehensive review of papers covering user perspectives, published before 2010 and was a main motivation of our study. Hence, this work focuses on papers published since 2010.
\end{itemize}

Based on the above findings, we decided to define the following more focused research questions for the SLR in the second stage of our work, as presented in Table~\ref{tab:RQ}.

\begin{table}[!htb]
\centering
\caption{Research questions (RQs) identified for the follow-up SLR via the meta-review}
\label{tab:RQ}
\begin{tabularx}{\linewidth}{lX}
\toprule
RQ1 & Has the current literature paid attention to the security and privacy perspectives of users within a home network in a holistic manner?\\
RQ2 & Is there any research exploring home users perspectives on security and privacy of multiple inter-connected devices in a home network?\\
RQ3 & What is the current research on users' awareness of security and privacy issues of a home network?\\
RQ4 & To what extent researchers have explored the security and privacy concerns of users in a home network?\\
RQ5 & What type of user security behaviors and practices in a home network have been researched?\\
RQ6 & What research has been conducted to understand the difference in users security and privacy behaviors and practices in a multi-user home network?\\
RQ7 & What demographic factors have been studied while exploring home users' perspectives on security and privacy aspects in a home network?\\
RQ8 & How have different contexts of a home network been considered when studying home users' perspectives regarding security and privacy aspects?\\
RQ9 & What theoretical and conceptual frameworks have been proposed and used to facilitate studies on user perspectives regarding security and privacy aspects of a home network?\\
\bottomrule
\end{tabularx}
\end{table}

As is evident from the paper, the boundary between RQ3, RQ4 and RQ5 is not a clear-cut one. Broadly speaking, RQ3 focuses more on the knowledge of home users on \textit{facts} related to security and privacy matters in the home networking context; RQ4 focuses more on home users' concerns (i.e., perceived risks and problems), including not only concerns caused by awareness of genuine security and privacy issues, but also those caused by false perception or misunderstanding of (possibly non-existing) security and privacy issues; and RQ5 looks more at what home users actually do (behaviors and practices) and the reasons behind them (attitudes, motivation, perception). RQ5 has a broader scope and may be argued to include RQ3 and RQ4 as two sub-questions, since actual behaviors and practices may also be caused by (lack of) awareness and/or concerns.

\section{Systematic Literature Review (SLR)}
\label{sec:SLR}

The SLR was conducted in line with the 9 research questions identified via the meta-review reported at the end of the previous section, examining related \emph{original} (i.e., non-review) research work in the literature from 2010 till 2021. There are two reasons why we decided to exclude papers published before 2010. First, we wanted to focus on more recent research (the past decade), which is a common practice for SLRs~\cite{Kaur2019SysReview, Uchendu2021SysReview}. Our results showed that relevant research was indeed more done in the past several years (see Figure~\ref{fig:PapersByYear}). Second,  \citeauthor{howe2012psychology}'s review work~\cite{howe2012psychology} conducted in \citeyear{howe2012psychology} covered original research papers published before 2010 quite comprehensively.

\subsection{Methodology}
\label{subsec:SLRMethodology}

As mentioned in Section~\ref{sec:GenMethod}, we followed the same PRISMA method~\cite{moher2009preferred, Page2021PRISMA_Update} for the SLR, following the exclusion and inclusion criteria explained below. Similar to what we did for the mata-review, we also considered the term ``home'' in a broader sense.

\subsubsection{Exclusion Criteria}
\label{subsubsec:SLR_Exclusion}

Papers meeting the following exclusion criteria were excluded:

\begin{itemize}
\item Papers published before 2010

\item Non-English papers

\item Papers that are not related to a home network context or do not include a significant coverage of home networks

\item Papers that do not cover any security or privacy aspects

\item Papers that do not cover home user issues

\item Papers that have been considered in the meta-review reported in the previous section, or papers that do not report original research work.
\end{itemize}

\subsubsection{Inclusion Criteria}
\label{subsubsec:SLR_Inclusion}

Any topics which meets at least one of the  relevant areas in the RQs listed in Table~\ref{tab:RQ}.

\subsubsection{Information Sources \& Search Strategy}

As mentioned for the general methodology in Section~\ref{sec:GenMethod}, the same three databases, i.e., Scopus, ACM Digital Library and IEEE Xplorer, were used to search for related research papers. The scope of our SLR was decided mainly by the result of the meta-review which are reflected in our research questions Table~\ref{tab:RQ}. The research questions which were formulated as the result of the meta-review guided the process of a keyword selection. The search queries we used for the SLR are formed by four \emph{required} sets of keywords each covering a different aspect of our RQs: the context of home network or smart home, security or privacy, users, and behaviors (see Table~\ref{tab:SLR_keywords}).

\begin{table}[!htb]
\centering
\caption{The keywords used in the search queries of the SLR}
\label{tab:SLR_keywords}
\begin{tabularx}{\linewidth}{c X}
\toprule
& (Home AND (Network* OR Smart OR Comput* OR Internet OR Device))\\
\midrule
AND & (Security OR Privacy)\\
\midrule
AND & (User OR Human OR People OR Customer OR Person)\\
\midrule
AND & (Perception OR Awareness OR Concern OR Behaviour OR Behavior OR Worry OR Action OR Decision)\\
\bottomrule
\end{tabularx}
\end{table}

\subsection{Results}
\label{subsec:SLR_results}

\subsubsection{Papers Selected}

The papers returned from the database searches totaled: 2,521 from Scopus, 61 from ACM Digital Library, and 424 from IEEE Xplore. After removing duplicates, those papers were first screened based on their titles and abstracts, by applying the exclusion criteria. This resulted in 226 papers eligible for more detailed screening and selection based on their full text. This led to the exclusion of 124 papers and 102 eligible papers were selected.

We performed an additional snowballing-based process~\cite{Ghaljaie2017Snowball} to identify more relevant papers by analyzing references of the 102 papers selected. Any potentially relevant papers identified went through the same exclusion and inclusion criteria. In total 24 additional papers were identified following this snowballing process, increasing the number of selected papers to 126. The results of the process of searching for and identifying the final selected papers are shown in Figure~\ref{fig:PRISMA}.

\subsubsection{Yearly Trend of Selected Papers}

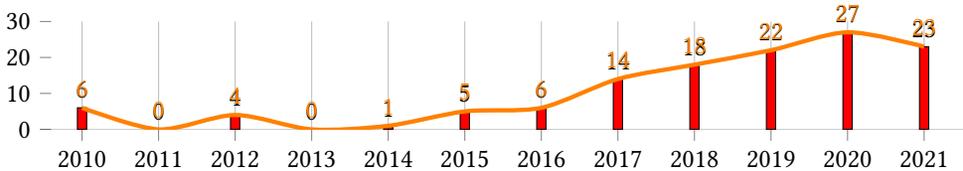
\begin{figure}[!htb]
\begin{tikzpicture}
\begin{axis}[
    width=\linewidth,
    height=3cm,
    ybar=0pt,
    bar width=0.25cm,
    enlarge x limits = {0.05},
    axis lines*=left,
    y axis line style={opacity=0},
    x axis line style={opacity=.2},
    ymin=0,
    ymax=30,
    xmajorgrids=true,
    nodes near coords,
    x tick label style={/pgf/number format/.cd, set thousands separator={}},
    xtick=data
    ]
\addplot [fill=red,ybar,bar width=3.5pt]
    coordinates { (2010,6)(2011,0)(2012,4)(2013,0)(2014,1)(2015,5)(2016,6)(2017,14)(2018,18)(2019,22)(2020,27)(2021,23)};
\addplot [ultra thick,orange,line join=round,smooth] 
    coordinates  { (2010,6)(2011,0)(2012,4)(2013,0)(2014,1)(2015,5)(2016,6)(2017,14)(2018,18)(2019,22)(2020,27)(2021,23)};
\end{axis}

\end{tikzpicture}%
\caption{The number of papers published yearly between 2010 and 2021.}
\label{fig:PapersByYear}
\Description{This chart demonstrates the increase in publication frequency of related papers after 2016 from 6 to 27 in 2020 and 22 in 2021.}
\end{figure}

Figure~\ref{fig:PapersByYear} demonstrates the yearly trend of selected papers in the past 11 years (2010--2021). As can be seen, the majority of papers were published after 2016, indicating the fact that related research has gained momentum towards the second half of the 2010s. This trend is not surprising given the fact that smart home devices have become more popular more recently, and privacy and security issues around them has become more prominent in the past a few years.

\subsubsection{Thematic Analysis of Selected Papers}
\label{subsubsec:SLR_thematic_analysis}

\begin{table}
\scriptsize
\caption{Papers in 9 RQs (2016--2021). Legend: \fullcirc\ Substantial coverage \halfcirc\ Partial coverage \qcirc\ Light touch}
\label{tab:papers2RQs_2016-20}
\setlength\extrarowheight{2pt}
\begin{NiceTabular}{cc*{9}{c}}[hvlines]

\textbf{Period} & \textbf{Paper(s)} & \textbf{RQ1} & \textbf{RQ2} & \textbf{RQ3} & \textbf{RQ4} & \textbf{RQ5} & \textbf{RQ6} & \textbf{RQ7} & \textbf{RQ8} & \textbf{RQ9}\\

\Block{49-1}{\rotate 2016--2021}

& \cite{Markey_2020_AllinOne_Behavior, sikder2019multi} & & \fullcirc & & & & & & &\\
& \cite{Ahmad_2020_TangibleMuser, he2018rethinking, zeng2017end} & & & & & \halfcirc & \fullcirc & &  &\\
& \cite{he2018rethinking} & & & & & \halfcirc & \fullcirc & & \halfcirc &\\
& \cite{apthorpe2018spying, mccreary2016context, Molina_2019_Contextual} & & & & & & & & \fullcirc &\\
& \cite{barbosa2020privacy} & \fullcirc & & & \halfcirc &\halfcirc & & & &\\
& \cite{binns2017Privacy} & & & \fullcirc & & \halfcirc & & & &\\
& \cite{javed2019alexa} & & &\fullcirc & \halfcirc & & & & &\\
& \cite{Yao2019PrivacyDesign} & & & \halfcirc & \halfcirc & & & & &\\
& \cite{boussard2018future} & \halfcirc & \fullcirc & & & & & & &\\

& \cite{bugeja2016privacy, nthala2018informal, schomakers2020understanding, sturgess2018capability, zhao2020unraveling} & \fullcirc & & & & & & & &\\
& \cite{cannizzaro2020trust, Bian2021Concerns} & & & & \halfcirc & & & \fullcirc & &\\
& \cite{chalhoub2020alexa, crabtree2017repacking} & & & & \fullcirc & \fullcirc & & & &\\
& \cite{Cobb_2021_MultiUser, Li_2020_Priviledge_Behavior,  Nthala_2017_IfitisUrgent_Practice, nthala2018rethinking, taieb2019user, topa2018usability, Lafontaine2021concerns} & & & & & \fullcirc & & & &\\
& \cite{dupuis2018help, Patterson2021Theory} & & & \fullcirc & & \fullcirc & & & & \fullcirc\\
& \cite{abdi2019more, ghiglieri2017context, kim2019innocent, mcdermott2019evaluating, ParkLim_2020_UserExpectations, Knutzen2021Awareness, Bermejo2021Awareness, Haney2021Behaviour, Schomakers2021Behaviour} & & & \fullcirc & & & & & &\\
& \cite{Ferraris_2020_TrustModel_HomeNetwork} & \fullcirc & & & & & & & & \halfcirc\\
& \cite{forget2016UserEngagement, kaaz2017understanding} & & & \qcirc & & \fullcirc & & & &\\
& \cite{freudenreich2020responding} & & & \fullcirc & & \halfcirc & & & &\\
& \cite{hwang2018consumers} & & & \halfcirc & \fullcirc & \qcirc & & & &\\
& \cite{Furini_2020_OntheUsage_Concern, nohlberg2020exploring, Seidi_2020_Please_Behavior, singh2018users, Heek_2017_Demography, Choukou2021Concerns} & & & & & \halfcirc & & \fullcirc & &\\
& \cite{al2015security, chhetri2019eliciting, fruchter2018consumer, Seymore_2020_Stranger_Concern, zimmermann2019assessing, Gruenewald_2020_TrustGap, Almutairi2021Concerns, Lutz2021Concerns, Sun2021Behaviour, Reinhardt2021Concerns} & & & & \fullcirc & & & & &\\
& \cite{geeng2019Multi_User, Jang2017Multi-user} & & & & & \qcirc & \fullcirc & & &\\
& \cite{guhr2020privacy} & & & & \fullcirc & & & & & \halfcirc\\
& \cite{huang2019perception} & & & & \fullcirc & & & \qcirc & &\\
& \cite{huang_2020_amazon} & & & & \fullcirc & & \fullcirc & & &\\
& \cite{bernd2019BystanderDomestic, bernd2020bystanders} & & & & \halfcirc & & \fullcirc & & &\\
& \cite{KIM_2019_Willingness, George2021Theory} & & & & & & & & & \fullcirc\\
& \cite{klobas2019perceived} & & & & & & & \fullcirc & & \fullcirc\\\cline{2-11}
& \cite{Kulyk_2020_Awareness, wickramasinghe2019survey, zheng2018user, Ahmad_2020_TangibleMuser} & & & \fullcirc & & \qcirc & & & &\\

& \cite{lee2020home} & & & & \fullcirc & & \fullcirc & & & \halfcirc\\
& \cite{lee2016context} & & & \halfcirc & \fullcirc & \qcirc & & & \fullcirc &\\
& \cite{lau2018alexa} & & & \qcirc & \halfcirc & \halfcirc & \fullcirc & & &\\
& \cite{liao2019understanding} & \fullcirc & & \halfcirc & \halfcirc & & & & &\\
& \cite{malkin2019privacy} &\fullcirc & &\halfcirc & \qcirc & \qcirc & \qcirc & & &\\
& \cite{manikonda2018s, psychoula2018users, golbeck2017user} & & & & \fullcirc & \halfcirc & & & &\\
& \cite{Park_2020_UserCognitive_Conceptual} & & & & \fullcirc & \qcirc & & & & \fullcirc\\
& \cite{Markey_2020_YouJust_MultiUser, Markey_2020_Idont_MultiUser} & & & \qcirc & & \qcirc & \fullcirc & & &\\
& \cite{mcgill2017old} & & & & & \halfcirc & & \qcirc & \fullcirc & \fullcirc\\
& \cite{mills2019empirical} & & & & & \halfcirc & & & & \fullcirc\\
& \cite{naeini2017privacy} & & & & \qcirc & \halfcirc & & & \fullcirc &\\
& \cite{pal2019embracing} & \fullcirc & & & \qcirc & \qcirc & & & &\\
& \cite{prasad2019understanding, worthy2016trust, Ghosh_2020_Understanding_Behavior, Mols2021Concerns, alam_2017_review, Mols2021Concerns, Alam2021Concerns} & & & & \fullcirc & \halfcirc & & & &\\
& \cite{reeder2020older} & & & & \qcirc & & & \fullcirc & &\\
& \cite{tabassum2019perception} & & & & & \qcirc & & & \fullcirc &\\
& \cite{thompson2017security, Tsai_2016_Understanding_Conceptual, white2017analysis, yang2017user} & & & & & \qcirc & & & & \fullcirc\\\cline{2-11}
& \cite{yao2019privacy, Marky2021multi-user} & & & \qcirc & & & \fullcirc & & &\\
& \cite{zeng2017end} & & & \halfcirc & \halfcirc & \qcirc & \fullcirc & & &\\
& \cite{Choukou2021Concerns} & & & & \fullcirc & \qcirc & & \halfcirc&  &\\
& \cite{Frick2021Theory, Pal2021Theory} & & & & \halfcirc & & & &  & \fullcirc\\

\Block{1-2}{\textbf{Number of Papers}} & & \textbf{11} & \textbf{5} & \textbf{34} & \textbf{46} & \textbf{63} & \textbf{16} &
\textbf{17} & \textbf{8} & \textbf{21}\\
\end{NiceTabular}
\end{table}

\begin{table}
\scriptsize
\caption{Categorization of papers into the 9 RQs (2010--2015). The same legend as in Table~\ref{tab:papers2RQs_2016-20}.}
\label{tab:papers2RQs_2010-15}
\setlength\extrarowheight{2pt}
\begin{NiceTabular}{cc*{9}{c}}[hvlines]

\textbf{Period} & \textbf{Paper(s)} & \textbf{RQ1} & \textbf{RQ2} & \textbf{RQ3} & \textbf{RQ4} & \textbf{RQ5} & \textbf{RQ6} & \textbf{RQ7} & \textbf{RQ8} & \textbf{RQ9}\\

\Block{7-1}{\rotate 2010--2015}

& \cite{al2015security} & & \fullcirc & & & & & & &\\
& \cite{Anderson2010practicing} & & & & & \halfcirc & & & & \fullcirc\\
& \cite{Jin_2012_Awareness, kang2015MentalModel, kritzinger2010Awareness, xavier2012Threat, talib2010analysis} & & & \fullcirc & & & & & &\\
& \cite{liang2010understanding} & & & & & & & & & \fullcirc\\
& \cite{Oulasvirta_2012_Long-term_Behavior, wash2015SecurityBelief, ho2010SecurityChoice, Jin_2012_Awareness} & & & & & \fullcirc & & \fullcirc & &\\
& \cite{Wilkowska_2012_Privacy_Demography} & & & & \fullcirc & & & \fullcirc& &\\
& \cite{wilkowska2015perceptions} & & & & & & & \fullcirc & &\\

\Block{1-2}{\textbf{Number of Papers}} & & \textbf{0} & \textbf{1} & \textbf{5} & \textbf{1} & \textbf{5} & \textbf{0} & \textbf{6} & \textbf{0} & \textbf{2}\\
\end{NiceTabular}
\end{table}

In order to answer the nine RQs defined for the SLR, we conducted a thematic analysis of all selected papers and classified them into nine topical themes each corresponding to an RQ, as shown in Table~\ref{tab:papers2RQs_2016-20} (2016--2021) and Table~\ref{tab:papers2RQs_2010-15} (2010--2015). Considering the overlaps between RQ3, RQ4 and RQ5, we mapped selected papers to them as follows: papers that explicitly refer to home users' understanding, awareness, perception, knowledge and belief on security and privacy matters in the context of a home network were categorized under RQ3; papers that have an explicit coverage on home users' concerns or worried on security and privacy issues and problems were categorized under RQ4; and papers that cover home users' actual security and privacy behaviors and practices were categorized under RQ5. Comparing papers published in the past six years (2016--2021) and the earlier six years (2010--2015), we can see two revealing patterns: 1) research on related topic has been increasing drastically recently since 2016; and 2) recent papers frequently cover multiple RQs compare to earlier papers, indicating that more researchers realized the complexity of security and privacy issues of home networks and the need to study them from multiple angles. We represented the inclusion of this complexity in the included papers by the following specific graphic symbols: (a) ``Substantial coverage'' \fullcirc\ -- to represent papers which provided a substantial coverage of one RQ, (b) ``Partial coverage'' \halfcirc\ -- when a part of the paper is devoted to one RQ, and finally (c) `` Light touch`` \qcirc\ -- where a paper discussed about an RQ only briefly. Discussions for each RQ mainly concentrated on the papers which provided a full coverage, while mentioning about the partial coverage. Across all papers published in the 11 years our SLR covered, we can see some popular research areas (e.g., RQ4 ``Privacy concerns'', RQ5 ``Security behaviors and practices'', and RQ9 ``Theoretical and conceptual frameworks''). In addition, the results also revealed some obviously less-studied areas, especially under RQ2 on multiple inter-connected devices and RQ8 on contextual aspects.

For each RQ, we will discuss the relevant papers covered in our SLR with greater details later, grouping them into different sub-categories within each RQ. Table~\ref{tab:RQ-subCategory} shows an overview of all the papers which have been mapped to the RQs and the sub-categories within each RQ.

\begin{table}[!ht]
\caption{A taxonomic view of the research papers under different sub-categories belonging to the RQs}
\label{tab:RQ-subCategory}
\scriptsize
\setlength\extrarowheight{2pt}
\begin{tabularx}{\textwidth}{|l|l|X|c|}
\hline
\textbf{\small{RQ}} & \textbf{\small{Sub-category}} & \textbf{\small{Paper(s)}} & \textbf{\small{\#}}\\
\hline

\multirow{5}{*}{\textbf{RQ1: Holistic View}}
& Smart device adoption & \cite{barbosa2020privacy, pal2019embracing, liao2019understanding} & 3\\
& General challenges \& solutions & \cite{bugeja2016privacy} & 1\\
& General benefits \& risks & \cite{malkin2019privacy, wilson2015smart} & 1\\
& Data flows \& privacy risks & \cite{sturgess2018capability} & 1\\
& The role of trust & \cite{Ferraris_2020_TrustModel_HomeNetwork, schomakers2020understanding} & 2\\
& Hybrid nature of modern smart home & \cite{zhao2020unraveling, Nthala_2017_IfitisUrgent_Practice, boussard2018future} & 3\\
\hline

\multirow{3}{*}{\textbf{RQ2: Multiple Devices}}
& Lack of common authentication techniques & \cite{al2015security} & 1\\
& Multi-device and multi-user scenarios & \cite{boussard2018future, sikder2019multi} & 2\\
& Multi-device privacy configuration & \cite{Markey_2020_AllinOne_Behavior} & 1\\
\hline

\multirow{4}{*}{\textbf{RQ3: User Awareness}}
& Effect of workplace training & \cite{talib2010analysis, kritzinger2010Awareness, kang2015MentalModel, mcdermott2019evaluating} & 4\\
& Privacy awareness \& perception & \cite{zheng2018user, dupuis2018help, ghiglieri2017context, Markey_2020_YouJust_MultiUser, Markey_2020_Idont_MultiUser, schomakers2020understanding, Patterson2021Theory, Marky2021multi-user, binns2017Privacy, Knutzen2021Awareness, Bermejo2021Awareness, wickramasinghe2019survey, zeng2017end, freudenreich2020responding} & 14\\
& Smart personal assistants & \cite{javed2019alexa, malkin2019privacy, abdi2019more, Kulyk_2020_Awareness, ParkLim_2020_UserExpectations} & 5\\
& Different types of home users & \cite{kim2019innocent, xavier2012Threat,ahmad2018cyber} & 3\\
& PrivSec awareness in general & \cite{kaaz2017understanding, forget2016UserEngagement, lee2016context, lau2018alexa, hwang2018consumers, Yao2019PrivacyDesign, yao2019privacy, liao2019understanding, malkin2019privacy} & 9\\
\hline

\multirow{6}{*}{\textbf{RQ4: User Concern}}
& Privacy concerns in general & \cite{zimmermann2019assessing, worthy2016trust, lee2020home, hwang2018consumers, psychoula2018users, Yao2019PrivacyDesign, Almutairi2021Concerns} & 7\\
& Concerns with smart speakers & \cite{lau2018alexa, malkin2019privacy, chalhoub2020alexa, Mols2021Concerns, Lutz2021Concerns, chhetri2019eliciting, javed2019alexa, fruchter2018consumer, manikonda2018s, huang_2020_amazon, Park_2020_UserCognitive_Conceptual} & 11\\
& Reflection on underlying behaviors & \cite{Ghosh_2020_Understanding_Behavior, liang2010understanding} & 2\\
& Parental privacy concerns & \cite{Alqhatani_2018_Exploring_COncern, prasad2019understanding, Sun2021Behaviour} & 3\\
& Other aspects & \cite{golbeck2017user, Seymore_2020_Stranger_Concern, Alam2021Concerns, Gruenewald_2020_TrustGap, lee2016context, guhr2020privacy, crabtree2017repacking, barbosa2020privacy, cannizzaro2020trust, huang2019perception, wilkowska2015perceptions, Choukou2021Concerns, Reinhardt2021Concerns} & 13\\
& PrivSec concern in general& \cite{bernd2020bystanders, bernd2019BystanderDomestic, pal2019embracing, yao2019privacy, naeini2017privacy, zeng2017end, kang2015MentalModel, reeder2020older, Alam2021Concerns, Frick2021Theory, Frick2021Theory} & 11\\
\hline

\multirow{7}{*}{\textbf{RQ5: Behavior/Practice}}
& Security practice related decision & \cite{Nthala_2017_IfitisUrgent_Practice, nthala2018informal, nthala2018rethinking, Alam2021Concerns} & 4\\
& Management and configuration & \cite{kaaz2017understanding, ho2010SecurityChoice, topa2018usability} & 3\\
& Security behavior affecting practice & \cite{Anderson2010practicing, he2018rethinking, binns2017Privacy, dupuis2018help, taieb2019user, worthy2016trust, freudenreich2020responding} & 7\\
& Personalized approaches & \cite{wash2015SecurityBelief, forget2016UserEngagement} & 2\\
& Privacy-specific behaviors & \cite{crabtree2017repacking, chalhoub2020alexa,  lau2018alexa, malkin2019privacy, Al-Ameen2021Behaviour} & 5\\
& Machine Learning based study & \cite{naeini2017privacy, barbosa2020privacy, manikonda2018s, Li_2020_Priviledge_Behavior} & 4\\
& Other aspects & \cite{prasad2019understanding, Jin_2012_Awareness, Oulasvirta_2012_Long-term_Behavior, Abrokwa2021Behaviour} & 4\\
& PrivSec behavior in general& \cite{Ahmad_2020_TangibleMuser, freudenreich2020responding, Kulyk_2020_Awareness, Furini_2020_OntheUsage_Concern, Park_2020_UserCognitive_Conceptual, Ghosh_2020_Understanding_Behavior, Seidi_2020_Please_Behavior, pal2019embracing, wickramasinghe2019survey, tabassum2019perception, zeng2019HomeUser, lau2018alexa, Markey_2020_YouJust_MultiUser, Markey_2020_Idont_MultiUser, geeng2019Multi_User, Jang2017Multi-user, mcgill2018gender, nohlberg2020exploring, Heek_2017_Demography, zheng2018user, manikonda2018s, psychoula2018users, singh2018users, golbeck2017user, naeini2017privacy, thompson2017security, lee2016context, yang2017user, white2017analysis, Tsai_2016_Understanding_Conceptual, mills2019empirical, Lafontaine2021concerns, Mols2021Concerns, Choukou2021Concerns, Patterson2021Theory} & 35\\
\hline

\multirow{2}{*}{\textbf{RQ6: Multi-User}}
& Bystander privacy & \cite{yao2019privacy, bernd2019BystanderDomestic, bernd2020bystanders, Ahmad_2020_TangibleMuser, Markey_2020_AllinOne_Behavior,Cobb_2021_MultiUser} & 6\\
& Access control \& configuration & \cite{Jang2017Multi-user, geeng2019Multi_User, zeng2019HomeUser, he2018rethinking, Markey_2020_AllinOne_Behavior} & 5\\
& Devices shared by multiple users & \cite{huang_2020_amazon, geeng2019Multi_User, Jang2017Multi-user, malkin2019privacy, lau2018alexa, Markey_2020_Idont_MultiUser, Marky2021multi-user} & 7\\
\hline

\multirow{3}{*}{\textbf{RQ7: Demography}}
& Gender & \cite{Wilkowska_2012_Privacy_Demography, nohlberg2020exploring, mcgill2018gender, Furini_2020_OntheUsage_Concern, lee2020home} & 5\\
& Non-gender demographic factors & \cite{Heek_2017_Demography, wilkowska2015perceptions, Choukou2021Concerns, reeder2020older, singh2018users, cannizzaro2020trust, wash2015SecurityBelief, klobas2019perceived, Bian2021Concerns}& 9\\
& Location & \cite{Seidi_2020_Please_Behavior, huang2019perception, Lafontaine2021concerns} & 3\\
\hline

\multirow{2}{*}{\textbf{RQ8: Contextual}}
& Location as context & \cite{mccreary2016context, Molina_2019_Contextual, naeini2017privacy} & 3\\
& Device as context & \cite{lee2016context, apthorpe2018spying, tabassum2019perception, mcgill2017old, he2018rethinking, Lutz2021Concerns} & 6\\
\hline

\multirow{4}{*}{\textbf{RQ9: Theoretical}}
& Protection Motivation Theory (PMT) & \cite{mills2019empirical, white2017analysis, mcgill2017old, dupuis2018help, Anderson2010practicing, George2021Theory} & 6\\
& Extended PMT & \cite{Tsai_2016_Understanding_Conceptual, thompson2017security} & 2\\
& Theory of Planned behavior (TPB) & \cite{yang2017user, guhr2020privacy, Ferraris_2020_TrustModel_HomeNetwork} & 3\\
& Conceptual Framework & \cite{liang2010understanding, pal2019embracing, KIM_2019_Willingness, Frick2021Theory, Pal2021Theory, Park_2020_UserCognitive_Conceptual, lee2020home} & 7\\
\hline
\end{tabularx}
\end{table}

\subsubsection{\textbf{RQ1: Holistic view}}
\label{subsubsec:RQ1}

For this RQ, we considered 11 papers that examined security and privacy of home networks from a more holistic point of view.

Papers in this category discussed different aspects, including privacy and security risks of users while adopting to smart devices at home~\cite{barbosa2020privacy, pal2019embracing, liao2019understanding}, general challenges for home users and solution~\cite{bugeja2016privacy}, comprehensively assessing privacy risks of smart home by investigating the data-collecting capabilities of its integral components and assessing the individual risk they pose~\cite{sturgess2018capability}, general benefits and risks~\cite{malkin2019privacy, wilson2015smart}, the role of trust~\cite{Ferraris_2020_TrustModel_HomeNetwork, schomakers2020understanding}, and hybrid nature of modern smart homes~\cite{zhao2020unraveling, Nthala_2017_IfitisUrgent_Practice, boussard2018future}. The last two aspects are of particular interest for RQ1, so we briefly introduce the five papers below with greater details.

\textbf{The role of trust}: \citeauthor{Ferraris_2020_TrustModel_HomeNetwork}~\cite{Ferraris_2020_TrustModel_HomeNetwork} explored the trust relationship between the user and popular smart devices and suggested an interesting trust model, which would enhance home security in regards to how devices interact with users and other devices. \citeauthor{schomakers2020understanding}~\cite{schomakers2020understanding} discussed how the degree of automation can affect the privacy and trust perception of smart home users by not only exploring privacy from the information security viewpoint but also reflecting on the physical, social and psychological dimensions.

\textbf{The hybrid nature of modern smart homes}: \citeauthor{zhao2020unraveling}~\cite{zhao2020unraveling} reflected on the legal definition for the current digital home which has a fluid boundary and discussed some rising problems such as the changing perception of the home's location, the increasing significance of data protection in the home, and the weakening legal enforcement owing to the `cross border data-flows' and `complicated industrial supply chain'. \citeauthor{Nthala_2017_IfitisUrgent_Practice}~\cite{Nthala_2017_IfitisUrgent_Practice} investigated data security decisions in the home and proposed to study them by considering the home space in three distinct areas: social, activity-based, and technological spaces. \citeauthor{boussard2018future}~\cite{boussard2018future} proposed the concepts of `vPlace' and `vSpace', to address the problem of proliferation of poorly secured IoT devices at home. `vPlace' is the collection of all registered resources in the home network and external devices owned by home users along with their dynamic states of connection to the network of connected or not connected to the network, and `vSpace' refers to virtual spaces or rooms oriented towards different contexts of family, work (from home) or visitors.

Some recommended solutions proposed in research falling into this category are interesting, such as calculating privacy risks from the data collection capabilities of a device and dividing the whole home network into physical and virtual places. However, we felt that research should take into account co-existence of different (both smart and traditional) devices in a home network, exploring whether people's security and privacy behaviors and actions are different in different types of home networks such as in a student accommodation, an Airbnb or a shared flat.

\subsubsection{\textbf{RQ2: Multiple inter-connected devices}}
\label{subsubsec:RQ2}

For this theme, we found only a small number of (four) papers~\cite{al2015security, boussard2018future, sikder2019multi, Markey_2020_AllinOne_Behavior}, which are briefly introduced below.

\citeauthor{al2015security}~\cite{al2015security} focused on the unavailability of a common authentication technique using multiple digital devices. \citeauthor{boussard2018future}~\cite{boussard2018future}'s vPlace and vSpace concepts, introduced previously for RQ1, utilized the software-defined network (SDN) technique to manage access control in a multi-user home network with multiple devices. \citeauthor{sikder2019multi}~\cite{sikder2019multi} discussed the complex and conflicting demands of multiple users in a multi-device home compared to a single-users environments and suggested a new access control system called Kratos to enhance awareness of related environment. \citeauthor{Markey_2020_AllinOne_Behavior}~\cite{Markey_2020_AllinOne_Behavior} (lab-based study, $n=15$) found that users would prefer to have detailed information about each device, a clear status communication, more dynamic and rule-based settings, and delegation options to adjust privacy settings in a multi-device setting.

The limited research on RQ2 suggests the need for research in the field of interconnections and interactions between multiple home devices, and any consequential security and privacy problems.

\subsubsection{\textbf{RQ3: Security and privacy awareness of home users}}
\label{subsubsec:RQ3}

35 papers covered in our SLR studied users' awareness and perception of security and privacy in the context of home networking, with 24 papers substantially covering this topic.

\textbf{The effect of workplace training} and the subsequent cyber awareness for home users is a topic that has received some attention. While some researchers investigated organizational, social and personal factors that can affect cyber security awareness of home users and attributed the cyber awareness of the home users to personal initiatives' knowledge, others such as \citeauthor{talib2010analysis}~\cite{talib2010analysis} and \citeauthor{kritzinger2010Awareness}~\cite{kritzinger2010Awareness} recommended organized security awareness programs at the workplace to help boost home users' cyber awareness. \citeauthor{kang2015MentalModel}~\cite{kang2015MentalModel} and \citeauthor{mcdermott2019evaluating}~\cite{mcdermott2019evaluating} suggested the absence of a direct relationship between people's technical background and their security or privacy awareness.

The topic of \textbf{low privacy awareness} is another popular topic discussed by many researchers~\cite{zheng2018user, dupuis2018help, ghiglieri2017context, Markey_2020_YouJust_MultiUser, Markey_2020_Idont_MultiUser, Schomakers2021Behaviour, Patterson2021Theory, Marky2021multi-user}, owing to reasons such as the absence of audio-visual inputs~\cite{zheng2018user} or a low level of self-efficacy~\cite{Patterson2021Theory}. On the positive side, it was also reported that home users are willing to engage in privacy protective mechanism, if relevant tools are easy to understand and cheap~\cite{dupuis2018help}. Through an online role-playing exercise, \citeauthor{binns2017Privacy}~\cite{binns2017Privacy} (online role playing scenario, $n=27$), found that home users' privacy-related decisions are heavily influenced by their pre-existing perceptions of and relationships with companies (more precisely, mobile app suppliers). They suggested privacy-aware tools that can help users to incorporate such pre-existing contextual factors into their privacy-related decisions. Two studies~\cite{Knutzen2021Awareness, Bermejo2021Awareness} suggested using augmented reality related solutions to make home users aware of privacy issues and to encourage them to make more informed decisions. \citeauthor{wickramasinghe2019survey}~\cite{wickramasinghe2019survey} (survey, $n=229$), observed that users had a lack of knowledge of sensitive data collected by smart objects and the thirds parties receiving such data. \citeauthor{zeng2017end}~\cite{zeng2017end} (interview, $n=15$), suggested how users' vulnerability depends on the level of their privacy knowledge, and pointed out a clear mismatch between the awareness and power of the owner/administrator of the smart home in comparison to other home users. \citeauthor{freudenreich2020responding}~\cite{freudenreich2020responding} (interview, $n=16$), studied security practices of home users related to Wi-Fi security and found that, although they were mostly aware of the vulnerabilities, they found it difficult to address these issues.

Due to the growing popularity of \textbf{smart speakers} (also called smart/intelligent personal assistants or voice assistants -- we will use the shorter term ``smart speaker'' hereinafter), some studies~\cite{javed2019alexa, malkin2019privacy, abdi2019more} looked into different security and privacy aspects of smart speakers. They found that, home users were generally aware of personal data being stored on smart speakers, but not with the service provider and in some cases with other third parties. Most users were not even aware that they could review or delete the stored data~\cite{malkin2019privacy}. The incomplete mental model on different privacy issues~\cite{abdi2019more} or lack of knowledge on the matter~\cite{Kulyk_2020_Awareness} leads to various privacy concerns. \citeauthor{ParkLim_2020_UserExpectations}'s research~\cite{ParkLim_2020_UserExpectations} demonstrated that at times, when users are more aware of their privacy, they expected their own personal space oriented features in the smart speakers.

Three papers looked at \textbf{different types of home users}. \citeauthor{kim2019innocent}~\cite{kim2019innocent} (trend analysis $n=23$; interview $n=20$; survey $n=188$) categorized users into different groups depending on the level of their cyber awareness. For example, `Innocent Irene' (extreme low level of awareness) or `Parental Patrick' (people with a family to protect). They suggested designing smart devices to suit to home users' cyber personality. \citeauthor{xavier2012Threat}~\cite{xavier2012Threat} (survey, $n=324$) pointed out how lack of awareness affects the ability to understand security threats whereas \citeauthor{ahmad2018cyber}~\cite{ahmad2018cyber} focused on lack of parental awareness on the topic of cyber threat towards children at home.

Home user awareness and perception were also addressed in nine papers~\cite{kaaz2017understanding, forget2016UserEngagement, lee2016context, lau2018alexa, hwang2018consumers, Yao2019PrivacyDesign, yao2019privacy, liao2019understanding, malkin2019privacy}, while focusing on other related concepts such as privacy and security concerns, privacy and security behaviors of single and multi-users and contextual security at home.

Research on RQ3 has a \textbf{noticeable focus on privacy awareness}, possibly because privacy issues of smart devices such as smart speakers are more visible and understandable by home users than (technical) security issues. The research on the effect of workforce training was a topic we did not expect, showing how home and work contexts can be connected. Discussions on user awareness on topics such as issues from connected home, existing tools and supports to mitigate issues arising, understanding of home user on legal and economical help available in case of security breach are surprisingly thin and important areas to focus on.

\subsubsection{\textbf{RQ4: Security and privacy concerns of home users}}
\label{subsubsec:RQ4}

RQ4 is one of the most discussed themes, with 47 papers covering a range of related topics, with 36 of them focusing majorly on this RQ. The other 11 papers~\cite{bernd2020bystanders, bernd2019BystanderDomestic, pal2019embracing, yao2019privacy, naeini2017privacy, zeng2017end, kang2015MentalModel, reeder2020older, Alam2021Concerns, Frick2021Theory, Pal2021Theory} focused more on topics in other RQs such as multi-user concept or user behaviors, so they will be covered elsewhere.

Seven papers looked at \textbf{privacy concerns in a more broader and systematic sense}. In order to understand the privacy concerns of home users, \citeauthor{zimmermann2019assessing}~\cite{zimmermann2019assessing} (interview $n=42$) categorized home user concerns under two types, `unrelated to attacks' such as dependency on a technology or loss of control to use, and `related to attacks' such as smart home data exposure and manipulation of device sensors. \citeauthor{worthy2016trust}~\cite{worthy2016trust} (experiment, $n=5$) collected daily use data from the participated homes with a specially developed IoT device, to find that that users tend to demand more control over the information collection process when they have less trust on the data controller and consumers. Through an empirical analysis involving 265 valid respondents, \citeauthor{lee2020home}~\cite{lee2020home} (survey, $n=300$) observed a positive association between different types of vulnerabilities (`Technological', `Legal' and `User') and IoT privacy concern, except provider vulnerabilities. \citeauthor{hwang2018consumers}~\cite{hwang2018consumers} (survey, $n=300$) revealed different levels of privacy risk perception for different types of home IoT services, venturing into a trade-off between privacy and functionalities. This trade-off is also re-iterated by a study conducted by \citeauthor{psychoula2018users}~\cite{psychoula2018users}(survey $n=231$; interview $n=41$). \citeauthor{Yao2019PrivacyDesign}~\cite{Yao2019PrivacyDesign} explored user-centered privacy design (Co-design study, $n=25$) to demonstrate home users' conceptualization of privacy control mechanisms. They identified six factors (Data Transparency \& Control', `Security', `safety', `Usability', `Contextual detection \& Personalizing', and `System Modality') to help design smart home privacy controls. A study~\cite{Almutairi2021Concerns} focused on understanding Saudi home users' privacy and security concerns of using smart devices, and revealed that 79.7\% of Saudi users were afraid of losing their data because of low awareness of related issues and the lack of governmental interventions.

Eleven papers discussed \textbf{privacy concerns on smart speakers}, the most popular topic under RQ4, and discussed it from many angles, i.e., \textbf{lack of privacy concerns by users, contextual privacy concerns, data collection behaviors of the manufacturers, the underlying behaviors such as trust and psychological factors}. \citeauthor{lau2018alexa}~\cite{lau2018alexa} conducted a diary study with 34 smart speaker user and non-users, to find that smart speaker users lacked knowledge on privacy risks whereas non-users lacked trust on the vendors. This was further highlighted by \citeauthor{malkin2019privacy}~\cite{malkin2019privacy} and \citeauthor{chalhoub2020alexa}~\cite{chalhoub2020alexa}, who commented on how users preferred comfort to privacy because of the low level of privacy concern and any desire to observe privacy behaviors is inhibited by the lack of user-friendly interface. Concerns about the privacy risks that smart speakers can track and listen to users' data were observed in two studies~\cite{javed2019alexa, chhetri2019eliciting}. \citeauthor{fruchter2018consumer}~\cite{fruchter2018consumer} used different NLP (natural language processing) tools to process 109,536 online user reviews on Amazon Echo and Google Home for the presence of specific security and privacy-oriented keywords. Only 2\% of the reviews using those keywords showed major concerns about data collection. \citeauthor{manikonda2018s}~\cite{manikonda2018s} examined online reviews to note users' positive outlook towards the smart speakers and a good level of privacy. \citeauthor{Mols2021Concerns}'s explorative study (survey, $n=325$; focus group, $n=35$) on the Dutch households' privacy concerns~\cite{Mols2021Concerns} provided a multi-dimensional understanding of users' concerns including surveillance, device security, day-to-day user behaviors and transparency of platforms. \citeauthor{Lutz2021Concerns}~\cite{Lutz2021Concerns} (survey, $n=325$) discussed privacy concerns related to smart speakers from a contextual perspective, suggesting that such concerns vary depending on the source. Two papers focused on users' concerns about the \textbf{data collection behaviors of smart speaker manufacturers}. \citeauthor{huang_2020_amazon}~\cite{huang_2020_amazon} investigated shared use of smart speakers (interviews, $n=26$), and observed that participants expressed privacy concerns about their housemates and visitors, and also about privacy-invasive data collection by speaker manufacturers. \citeauthor{Park_2020_UserCognitive_Conceptual}~\cite{Park_2020_UserCognitive_Conceptual} (survey, $n=359$) noticed that privacy concerns led to negative privacy-coping behaviors such as anger, anxiety and disappointment against the relevant companies, and generate bad words of mouth or disengagement.

\textbf{Reflection on the underlying behaviors} such as psychological and trust behind users' concerns about smart speakers was explored in two papers. \citeauthor{liao2019understanding}'s investigation (survey, $n=1160$) on the role of privacy and trust in users' decisions on adopting smart speakers~\cite{liao2019understanding} revealed that, users tend to trust the vendors (Google, Amazon, and Apple) on the usage of their information, and privacy concerns differ between people who use different hardware devices (a smart phone or a smart speaker) to interact with the agent. \citeauthor{Ghosh_2020_Understanding_Behavior}~\cite{Ghosh_2020_Understanding_Behavior} (survey, $n=289$) examined different psychological mechanisms underlying home user's interaction with software agents such as Alexa and Siri and hardware devices such as smart speakers and smart phones, and how they affect privacy concerns and information disclosure behaviors. Similar to what \citeauthor{Ghosh_2020_Understanding_Behavior}~\cite{Ghosh_2020_Understanding_Behavior}'s study explained about how participants are more likely to report on higher privacy concerns when interacting with voice assistants through smart speakers than through smart phones.

\textbf{Parental privacy concern} was discussed in three papers. Reflecting on parent's technology usage, control and concerns, \citeauthor{Alqhatani_2018_Exploring_COncern}~\cite{Alqhatani_2018_Exploring_COncern} (interview, $n=20$) stated that parents did discuss children's privacy and security concerns regarding their online use and controls but did not expand on these to include smart devices. In contrast, \citeauthor{prasad2019understanding}~\cite{prasad2019understanding} (interview, $n=20$) considered parental privacy concerns towards service providers and manufacturers and how they affect their children. \citeauthor{Sun2021Behaviour}~\cite{Sun2021Behaviour} (interview, $n=23$) identified six factors which, according to them, influence parents' perception of privacy risks, including parenting style, tech-savviness, trust in manufacturer, age of the children, features of the used devices and news media reporting.

While twelve papers considered \textbf{other aspects which might affect privacy concerns} of home users. \citeauthor{golbeck2017user}~\cite{golbeck2017user} analyzed 501 online comments to find out that 81\% of the home user have concerns about using their Internet Service Providers (ISP) supplied home router as a public Wi-Fi hot-spot. Privacy concern was found to be the most prevalent ethical concern in a study (survey, $n=631$) by \citeauthor{Seymore_2020_Stranger_Concern}~\cite{Seymore_2020_Stranger_Concern}. Through their study on authentication management techniques of home users ($n=93$), \citeauthor{Alam2021Concerns}~\cite{Alam2021Concerns} revealed that this type of privacy concerns does not actually reflect users' practices. \citeauthor{Gruenewald_2020_TrustGap}~\cite{Gruenewald_2020_TrustGap}'s study (survey, $n=701$) revealed that participants were more inclined to share their location data for services (50\%), than with service providers (28\%), but struggled with differentiating between the two. \citeauthor{lee2016context}~\cite{lee2016context} conducted a clustering analysis based on data collected from 200 participants on  2,800 hypothetical IoT scenarios and five contextual parameters (`where', `what', `who', `reason' and `persistence') affecting home users' privacy concerns, to find scenarios according to their privacy risks. A new theoretical model (which will be discussed more in RQ9) was suggested by~\citeauthor{guhr2020privacy}~\cite{guhr2020privacy} to find out the role, privacy concerns plays in home users' acceptance of smart home technology. Results of \citeauthor{crabtree2017repacking}'s research~\cite{crabtree2017repacking} demonstrated the fact that the discussion and design efforts for any sort of privacy mechanisms should focus on managing human relationships rather than controlling data flows. Both \citeauthor{barbosa2020privacy}~\cite{barbosa2020privacy} and \citeauthor{cannizzaro2020trust}~\cite{cannizzaro2020trust} commented on users' reluctance on home IoT adoption due to their privacy concerns, and how such reluctance grows with age~\cite{cannizzaro2020trust}. Devices and platforms other than smart devices such as ambient assisted living (AAL)~\cite{Choukou2021Concerns}, mobile-assisted robots~\cite{Reinhardt2021Concerns} were also studied to understand the type of privacy concerns they might generate. \citeauthor{wilkowska2015perceptions}~\cite{wilkowska2015perceptions} used three different methodologies, i.e., a focus group($n=42$), a survey($n=104$) and an experimental usability study($n=55$), to study home users' privacy concerns in the context of AAL and found that the level of privacy concerns of participants observed, following each methodology differed significantly.

Finally, one paper paid attention to the role of cultural background in privacy concerns: \citeauthor{huang2019perception}~\cite{huang2019perception} conducted a small interview study with nine Chinese smart home users, showing some preliminary evidence regarding their privacy concerns likely being less than an average American user (based on past American privacy-related studies).

Multiple papers related to RQ4 \textbf{focused on a specific type of smart devices -- smart speakers}. The trade-offs between privacy and functionality, how privacy concerns affected the decision of smart home adoption, are also relatively popular topics of discussion. However, we found that \textbf{some smart home devices such as smart home appliances} are much less studied, despite the increasing use of these devices in modern smart homes.

\subsubsection{\textbf{RQ5: Security \& privacy behaviors and practices of home users}}
\label{subsubsec:RQ5}

This is another popular area of research, with 66 papers. Some papers (32) focused majorly on this RQ and others (34) reflected on security and privacy behaviors while discussing other concepts such as privacy concerns, awareness, multi-users issues and related conceptual frameworks~\cite{Ahmad_2020_TangibleMuser, freudenreich2020responding, Kulyk_2020_Awareness, Furini_2020_OntheUsage_Concern, Park_2020_UserCognitive_Conceptual, Ghosh_2020_Understanding_Behavior, Seidi_2020_Please_Behavior, pal2019embracing, wickramasinghe2019survey,tabassum2019perception, zeng2019HomeUser, lau2018alexa, Markey_2020_YouJust_MultiUser, Markey_2020_Idont_MultiUser, geeng2019Multi_User, Jang2017Multi-user, mcgill2018gender, nohlberg2020exploring, Heek_2017_Demography,zheng2018user, manikonda2018s, psychoula2018users, singh2018users, golbeck2017user, naeini2017privacy, thompson2017security, lee2016context, yang2017user, white2017analysis, Tsai_2016_Understanding_Conceptual, mills2019empirical, Lafontaine2021concerns, Mols2021Concerns, Choukou2021Concerns, Patterson2021Theory}. Here, we will focus on the 32 papers in the former group.

\textbf{Security practice related decision} was discussed in four papers. \citeauthor{Nthala_2017_IfitisUrgent_Practice}~\cite{Nthala_2017_IfitisUrgent_Practice} in their study, (interview, $n=15$) identified four themes ( stimuli (cues to action), support, stakeholders, and context) around home users' security practice related decisions. In a follow-up work (interviews, $n=50$) ~\cite{nthala2018rethinking}, they found that home users majorly rely on family and friends as an informal support network. Furthermore, in another follow-up work (survey, $n=1,128$; interview, $n=65$)~\cite{nthala2018informal}, they found that security practice is affected by survival/outcome bias, factors undermining confidence in a security measure, in addition to other well known factors such as trust, cost, knowledge and skill. \citeauthor{Alam2021Concerns}~\cite{Alam2021Concerns} observed that despite being very concerned about their privacy and security, users don't follow appropriate security steps.

\textbf{Management and configuration of smart devices} was explored in three papers. \citeauthor{kaaz2017understanding}~\cite{kaaz2017understanding} (experiment, $n=7$) found that, contrary to the popular belief, smart devices are not `plug and play' and a majority of home users face multiple barriers while configuring such devices, which often force them to accept vendors' default (possibly flawed) security and privacy settings. \citeauthor{ho2010SecurityChoice}~\cite{ho2010SecurityChoice}, in an earlier (\citeyear{ho2010SecurityChoice}) study on home wireless networks, revealed how home users depended on out-of-box security tools provided by manufacturers, hardly changed the default security settings, rarely installed and maintained encryption keys across devices, and did not perceive any differences between encryption and access control. \citeauthor{topa2018usability}~\cite{topa2018usability} collected and analyzed a broad spectrum of privacy and security issues on usability of security tools such as VPNs, anti-virus and anti-spyware programs (scenario-based, survey, $n=150$; interview, $n=112$) to highlight the need of detailed help, consistency regarding use of technical terms, concerns over the use of personal data, and the absence of appropriate usability tools for disabled users.

Seven studies looked at \textbf{security behaviors affecting home users' security practices} for home networks. \citeauthor{Anderson2010practicing}~\cite{Anderson2010practicing} observed (survey, $n=600$) that home users' security behaviors were influenced by cognitive, social and psychological components and hence, all such factors should be considered when analyzing their security attitudes. \citeauthor{he2018rethinking}~\cite{he2018rethinking} (survey, $n=425$) confirmed a proposed hypothesis that, participants focused on IoT devices' capabilities rather than the devices themselves to define access control and authentication policies. \citeauthor{binns2017Privacy}~\cite{binns2017Privacy} concluded that, home users' security actions are mostly influenced by their preconceived notion about the company responsible for the specific hardware device or software (e.g., a mobile app). However, \citeauthor{dupuis2018help}~\cite{dupuis2018help} were of the opinion that home users would be more willing to take precautionary actions if they clearly understand the security mechanisms of the products they use. Via a mixed-method study (survey, $n=1,006$; interview, $n=14$), \citeauthor{taieb2019user}~\cite{taieb2019user} observed that home users' attitudes and perception towards IoT devices and the security of their data depend on the device type, e.g., a smart health device could induce them to share more information than a smart speaker. \citeauthor{worthy2016trust}~\cite{worthy2016trust} conducted an experiment to collect data from five different households for a period of 10-14 days and analyzed the inhabitants' behaviors. They observed that the participants became familiar with new devices very soon and were generally happy with sharing data as long as it is properly de-identified (e.g., via aggregation) and they were aware of the purpose of collection. Risks associated with Wi-Fi vulnerabilities in home users were explored by \citeauthor{freudenreich2020responding}~\cite{freudenreich2020responding} (survey and interview, $n=16$), who found that, although generally aware of the privacy risks, people were not knowledgeable or skilled to address such problems.

\textbf{The need of personalized approaches} is highlighted in some work. Both \citeauthor{wash2015SecurityBelief}~\cite{wash2015SecurityBelief} and \citeauthor{forget2016UserEngagement}~\cite{forget2016UserEngagement} argued on the unsuitability of `one size fits all' approach to security. \citeauthor{wash2015SecurityBelief}~\cite{wash2015SecurityBelief} discussed different `folk models' of threats that induce a home user to make different security decisions according to their contextual belief. Whereas~\citeauthor{forget2016UserEngagement}~\cite{forget2016UserEngagement} collected user security behaviors and machine configurations of 73 users for at least 3 months within 9-month time window, and then interviewed 15 users out of the 73 ones. Their results reiterated a finding of \citeauthor{wash2015SecurityBelief}~\cite{wash2015SecurityBelief} -- since home users engage in security behaviors and practices differently, they may benefit from different styles and different levels of interventions for their specific needs.

\textbf{privacy-specific behaviors} of home users were explored in several papers. \citeauthor{crabtree2017repacking}~\cite{crabtree2017repacking} studied digital practices of 20 homes in the UK and found that participants employs a number of `fine-grained methods'(throwaway emails, ad blockers, cookies, consent forms, private browsing) to manage the flow of their private data securely. \citeauthor{chalhoub2020alexa}~\cite{chalhoub2020alexa} investigated home users' attitude in terms of the user experience (UX), and pointed out that the lack of users' privacy concerns arose out of their individual perception of the situation and how they traded their privacy needs for the benefits from smart devices.  \citeauthor{lau2018alexa}~\cite{lau2018alexa} and \citeauthor{malkin2019privacy}~\cite{malkin2019privacy} found no evidence of privacy-seeking behaviors in users of smart speakers, and observed that users did not use privacy controls already available to them in such devices. \citeauthor{Al-Ameen2021Behaviour}~\cite{Al-Ameen2021Behaviour} revealed a number of mismatches between users' actual perception of data collection and data sharing by the IoT devices compared to the devices' published privacy policies.

Four papers used \textbf{machine learning(ML) tools} to reflect on home users' privacy or security behaviors. \citeauthor{naeini2017privacy}~\cite{naeini2017privacy} used ML classifiers to predict users' preferred comfort level and their decision to allow or deny specific data collection, whereas \citeauthor{barbosa2020privacy}~\cite{barbosa2020privacy} implemented a decision tree classifier to suggest how easy affordability as a `motivator' can defocus privacy as a `blocker'. \citeauthor{manikonda2018s}~\cite{manikonda2018s} applied the latent Dirichlet allocation (LDA) algorithm~\cite{blei2003latent} and Word2Vec~\cite{mikolov2017advances} to understand users' privacy behavior and concerns. \citeauthor{Li_2020_Priviledge_Behavior}~\cite{Li_2020_Priviledge_Behavior} analyzed home users' security and privacy behavior by examining 15.4 million video streams from 211k Chinese users and observed that frequent use of the camera increases the privacy risks.

Four papers looked at \textbf{other aspects} of user attitudes, perceptions and behaviors.  \citeauthor{prasad2019understanding}~\cite{prasad2019understanding} discovered (focus groups, $n=3$; interviews, $n=14$) that parents did not trust device/software manufacturers or ISPs to protect their children from harms when using smart devices, and felt it was their responsibility to do so. \citeauthor{Jin_2012_Awareness}~\cite{Jin_2012_Awareness} discussed a different topic of residential privacy by examining data collected from the Foursquare application and identified several vulnerabilities and privacy risks caused by user behaviors. \citeauthor{Oulasvirta_2012_Long-term_Behavior}~\cite{Oulasvirta_2012_Long-term_Behavior} used a specific behavioral observation system (BOB) to pool sensor data from designated surveillance devices (Wi-Fi cameras, key-presses from personal computers, smart devices, TV and DVD media centers) to comment on the contextual behavioral change. \citeauthor{Abrokwa2021Behaviour}~\cite{Abrokwa2021Behaviour} (survey, $n=493$) observed no significant privacy behavioral differences between users of two mobile operating systems (iOS and Android).

The RQ5-related papers cover a wide range of topics, e.g., home users' privacy and security behaviors, practices and decisions, and parental behaviors, which mostly focused on issues around standalone devices. As in the case of RQ3 and RQ4, more future research should focus on understanding \textbf{home users' behaviors and attitudes in a connected home with hybrid devices}.

\subsubsection{\textbf{RQ6: Multiple users in a single home}}
\label{subsubsec:RQ6}

Eighteen papers in our study covered security and privacy issues in a multi-user home environment, in different levels of depth.

Five papers examined \textbf{privacy of bystanders}, e.g., guests and nannies. \citeauthor{yao2019privacy}~\cite{yao2019privacy} examined potential privacy concerns and expectations of the bystanders (Focus groups, $n=18$) and observed strong contextual variations, as they switch their roles under different social relationships. \citeauthor{bernd2019BystanderDomestic}~\cite{bernd2019BystanderDomestic} identified different types of bystanders including nannies, home care attendants, house cleaners and maintenance workers who can be affected by the use of smart devices but are not directly involved in the use of such devices. In a follow-up study, \citeauthor{bernd2020bystanders}~\cite{bernd2020bystanders} (interview, $n=25$) found that nannies expected the existence of smart cameras but wanted transparency of information from their employer (i.e., the homeowner) beforehand and were concerned by potential misuse of collected data. \citeauthor{Cobb_2021_MultiUser}~\cite{Cobb_2021_MultiUser} surveyed 386 incidental users of smart devices, to understand their most typical concerns and the context where it materializes, and recommended better communication between the primary and incidental users. \citeauthor{Ahmad_2020_TangibleMuser}~\cite{Ahmad_2020_TangibleMuser} (interview, $n=19$), proposed a concept of `tangible privacy' for designing IoT devices, to provide stronger privacy assurances to bystanders. 

Four other papers discussed challenges in a multi-user home, focusing on \textbf{access control and configuration managements of home devices by different users} in the same home environment. Access control issues in a multi-user home were discussed by three groups of researchers~\cite{Jang2017Multi-user, zeng2019HomeUser, he2018rethinking}, who conducted scenario-based analyses and discussed challenges of using smart devices in a multi-user environment, including coarse-grained access control resulting in either complete access or no access to users, intended or unintended threats to the primary user's data. \citeauthor{Markey_2020_AllinOne_Behavior}~\cite{Markey_2020_AllinOne_Behavior} investigated a prototype for multi-setting interface to adjust privacy settings by multiple users with multiple devices (experiment, $n=15$) and found that users prefer ability to access detailed information with the settings. By exploring security and privacy implications in a multi-user home, \citeauthor{zeng2019HomeUser}~\cite{zeng2019HomeUser} designed a prototype with different access control features (i.e., location-based, supervisory). They discovered that factors such as usability and configuration complexity, lack of concerns for devices, interference with other functionalities, trust between different home users are some of the reasons why users ignore access control mechanisms. Four researchers \cite{Jang2017Multi-user, he2018rethinking, zeng2019HomeUser} recommended different design changes in smart devices to increase the usability and accessibility of the functionalities to all home users. Additionally, \citeauthor{he2018rethinking}~\cite{he2018rethinking} exploring different types of relationship in a multi-user home (`Babysitter vs.\ visiting family', and `Child vs.\ teenager') found clear differences in different users' desires to have specific access control policies attached to different capabilities of an IoT device. 

Nine researchers discussed the \textbf{challenges of device use in a multi-user home by primary and secondary users}, mostly focusing on the shared use of smart speaker. Both \citeauthor{malkin2019privacy}~\cite{malkin2019privacy} and \citeauthor{lau2018alexa}~\cite{lau2018alexa} reflected on privacy tensions between primary, secondary and incidental users of smart speakers, while \citeauthor{zeng2017end}~\cite{zeng2017end} pointed out unique privacy and security challenges that occur in a multi-user home where incidental users depend on the primary users' knowledge and control.In a related, study (interview, $n=21$),\citeauthor{Markey_2020_YouJust_MultiUser}~\cite{Markey_2020_YouJust_MultiUser} observed that visitors would usually accept the data collection by smart devices so long as the data is anonymized and recommended to gain awareness and knowledge and evaluate data sensitivity, to exert control over their privacy. 
In a follow-up study, \citeauthor{Markey_2020_Idont_MultiUser}~\cite{Markey_2020_Idont_MultiUser} (interview, $n=42$) investigated two related privacy issues -- privacy of bystanders to homeowners and privacy of homeowners to bystanders. They recommended that, the IoT designers must pay attention to both bystanders and the users while designing their devices. In their study on mental model of 30 participants, \citeauthor{Marky2021multi-user}~\cite{Marky2021multi-user} noticed a general lack of awareness amongst visitors about the data flows in a smart home ecosystem. Based on interview data from 26 participants using smart speakers, \citeauthor{huang_2020_amazon}~\cite{huang_2020_amazon} (interview, $n=26$) found that participants had different types of concerns about inappropriate access and misuse of personal information by housemates and other external entities, but would follow the same risk management strategies in both cases. \citeauthor{geeng2019Multi_User}~\cite{geeng2019Multi_User} conducted a mixed-method study ($n=18$) to study the inter-communication, tensions, and challenges in a multi-user home. They observed that the smart home environment mimics the existing power dynamics (i.e., parent-child) in a household, giving smart home drivers more access to functionalities than other users. \citeauthor{ParkLim_2020_UserExpectations}~\cite{ParkLim_2020_UserExpectations} discussed on the privacy awareness of family members while sharing a smart speaker.  

Papers related to RQ6 cover mainly two areas, bystander privacy and access control. \textbf{Users' understanding of the more complicated data flows in a multi-user home} is one of the main topic that we felt should be studied more, along with other areas such as \textbf{threats from malicious secondary users} and \textbf{security and privacy implications of interactions between multiple devices and multiple users}.

\subsubsection{\textbf{RQ7: Demographic factors and their effects}}
\label{subsubsec:RQ7}

Seventeen papers in our study substantially covered topics related to this RQ.

Five papers studied home users' security behaviors, practices and concerns, with a focus on \textbf{gender}. \citeauthor{Wilkowska_2012_Privacy_Demography}~\cite{Wilkowska_2012_Privacy_Demography} studied the use of e-health technologies at home ($n=104$, 60 females and 44 males), showing that female and healthy adults were more prone to demand stringent security and privacy standards than male adults and ailing elderly, respectively. In contrast, \citeauthor{nohlberg2020exploring}'s survey (152 participants, 53 females and 99 males)~\cite{nohlberg2020exploring} found men to be more decisive in comparison to women in an information security decision. \citeauthor{mcgill2018gender}~\cite{mcgill2018gender} surveyed 624 users (234 females and 390 males) and their results echoed \citeauthor{nohlberg2020exploring}'s finding that security behaviors of female users are weaker than male users'. \citeauthor{Furini_2020_OntheUsage_Concern}~\cite{Furini_2020_OntheUsage_Concern} conducted a small study during the 2020 COVID-19 lockdown on people's privacy behaviors and concerns on smart speakers, and found that both male and female users had privacy concerns. Looking into gender and IoT use experience, \citeauthor{lee2020home}~\cite{lee2020home} concluded that female users were more concerned by their own vulnerabilities and people without technical experience were more concerned by providers' vulnerabilities.

Some other papers focused on \textbf{demographic factors beyond gender} such as age and disability while analyzing privacy and security attitudes and concerns of people using AAL, older adults and other variables such as age, ethnicity, income level and disability. \citeauthor{Heek_2017_Demography}~\cite{Heek_2017_Demography} studied the acceptance of AAL technologies, and the trade-offs between perceived benefits and barriers (survey, $n=279$). They found that user diversity in terms of age, disability and care giving experience does significantly affect the trade-offs between perceived benefits and barriers. In another related study that used three different methodological approaches (focus groups, survey and a usability study), \citeauthor{wilkowska2015perceptions}~\cite{wilkowska2015perceptions} noticed  privacy with regard to AAL technologies is independent of the age or gender. The same conclusion was also drawn by \citeauthor{Choukou2021Concerns}~\cite{Choukou2021Concerns}, who compared the attitude of older and younger adults on the use of AAL technology. However, when age was discussed as a factor influencing privacy and security attitude in general (not specific to AAL), it tends to play a significant role in user attitude. Age was a parameter studied by \citeauthor{singh2018users}~\cite{singh2018users} (survey, $n=231$) who found that older adults (36 -- 70) were more willing to share data on health grounds than their younger (below 36) counterparts. In their analysis of a large-scale survey with 2,033 UK participants, \citeauthor{cannizzaro2020trust}~\cite{cannizzaro2020trust} noticed that age and education level play a significant role in determining people's trust on IoT devices for security and privacy. In another study, \citeauthor{wash2015SecurityBelief}~\cite{wash2015SecurityBelief} found that educated users and older adults often exercise fewer precautions in regard to security threat. \citeauthor{klobas2019perceived}\cite{klobas2019perceived} reported similar results about security perceptions of older and educated participants who seemed to be more likely to assess security risks of IoT products and had a more positive attitude towards them, although they were still concerned about the privacy and functionality, which does not meet the need of their specific requirements.~\cite{Bian2021Concerns}. \citeauthor{reeder2020older}~\cite{reeder2020older} interviewed ten post-menopausal women (age band: 50-70) to understand their perception of wearable devices and found that the participants largely accepted the technology as useful.

\citeauthor{Seidi_2020_Please_Behavior}~\cite{Seidi_2020_Please_Behavior}'s study (survey, $n=214$) found that geo-privacy behaviors are very much linked to a participant's underlying knowledge of the field and similar across different demographic factors including gender, ethnicity and income level. \citeauthor{huang2019perception}~\cite{huang2019perception}'s comparative study on Chinese and US users revealed that users' privacy concerns in China seemed to be lower than in Western countries. \citeauthor{Lafontaine2021concerns}~\cite{Lafontaine2021concerns} conducted a survey ($n=232$) over three geographic regions (the US, the EU and India) and found that IoT users were comparatively comfortable in accepting risk than non-IoT users. Furthermore, they observed contrasting behaviors of users in different regions, i.e., people in India trust their government more in protecting their data compared to people in the US and the EU. 

Papers related to RQ7 studied demographic factors such as gender, age, educational background, disability, ethnicity and income level, and their implications on security and privacy behaviors of home users. There is \textbf{contradictory evidence of weaker security behaviors of females compared to males}, therefore needing more research in this area. We noticed that a majority of the studies focused on \textbf{developed nations such as the UK, the US and the EU}, so more research on developing countries and non-Western countries is much needed.

\subsubsection{\textbf{RQ8: Contextual factors influencing security and privacy behaviors and practices at home}}
\label{subsubsec:RQ8}

Eleven papers in our study contributed substantially to this theme.

Three papers considered \textbf{location} as the contextual factor. \citeauthor{mccreary2016context}~\cite{mccreary2016context}'s study (video experiment, $n=264$) found that people were very much concerned about privacy inside their home regardless of the activities they are involved in compared to outside their home. \citeauthor{Molina_2019_Contextual}~\cite{Molina_2019_Contextual} (survey, $n=276$) asked their participants to imagine using Wi-Fi networks at four different locations (coffee shop, university, Airbnb, and home). One of their hypotheses is that a higher belief in publicness heuristic can lead to less information disclosure. Their results showed positive evidence to support this hypothesis. \citeauthor{naeini2017privacy}~\cite{naeini2017privacy}'s research (vignette study, $n=1,007$) working with a set of 380 IoT data collection and different scenarios revealed the context-dependence and the diverse nature of privacy preferences of home users.

Other researchers analyzed \textbf{contextual use of devices}~\cite{Oulasvirta_2012_Long-term_Behavior} affecting privacy and security behaviors or leading to specific individual knowledge and experience or device's primary function dictating users' privacy perception~\cite{Ahmad_2020_TangibleMuser}. \citeauthor{lee2016context}~\cite{lee2016context} conducted a comprehensive analysis using K-mode clustering analysis, employing data from hypothetical IoT scenarios (survey, $n=200$). They used K-modes clustering analysis with four clusters (`Very unacceptable', `Unacceptable', `Somewhat acceptable', and `Acceptable') and analyzed the data with five contextual parameters (`where', `what', `who', `reason' and `persistence') to reflect on the impacts of contextual factors on peoples' privacy perceptions. \citeauthor{apthorpe2018spying}~\cite{apthorpe2018spying} used the contextual integrity (CI) framework (survey, $n=1,731$). They collected 3,849 information flows passing between various first and third-party recipients in a smart home to provide rich insights into why device manufacturers should survey privacy norms in specific contexts and why privacy norms should support restrictive rather than permissive IoT device communications. \citeauthor{tabassum2019perception}'s~\cite{tabassum2019perception} study, (interview, $n=23$) discovered that users' threat modelling of their home and their protection behaviors were not shaped by their existing knowledge, but by their experience in other computing contexts. \citeauthor{mcgill2017old}'s research (survey, $n=629$) on users' security behaviors~\cite{mcgill2017old} revealed that users perceived to expect more risks from the use of a mobile device than from a home computer. \citeauthor{he2018rethinking}~\cite{he2018rethinking} studied access control issues in multi-user homes ($n=425$) and looked at frequent context-dependent capabilities of various IoT devices. They noticed five contextual factors (`Age', `Location of Device', `Recent Usage History', `Time of Day' and `Location of User') impacted significantly on the implementation of access control capabilities of smart devices. According to \citeauthor{Lutz2021Concerns}'s study~\cite{Lutz2021Concerns}, privacy concerns are dependent on the context of the origin source.

The major areas of discussion on this RQ were \textbf{location} and \textbf{use of smart devices}. However, we feel that some other important contexts need more research, including varying security behaviors and practices of home users when using traditional devices verses smart devices, sharing a particular device with other users, different types of smart devices other than more studied ones such as smart speakers. There is also a lack of research on the legal context, e.g., home users' understanding of the legal support available to them in case of any security or privacy breaches, and their legal rights when it comes to the storage and manipulation of their data on different types of home devices.

\subsubsection{\textbf{RQ9: Theoretical frameworks of security and privacy behaviors}}
\label{subsubsec:RQ9results}

With Eighteen papers in our data set, RQ9 is another much-discussed theme in our study.

Seven papers in this theme~\cite{ mills2019empirical, white2017analysis, mcgill2017old, dupuis2018help, Anderson2010practicing,klobas2019perceived, George2021Theory} discussed users' security behaviors in light of the \textbf{PMT (Protection Motivation Theory)\cite{rogers1975protection}}.  Analyzing the survey data from 72 home computer users, \citeauthor{mills2019empirical}~\cite{mills2019empirical} concluded that participants were not significantly influenced by perceived vulnerability or perceived severity when trying to implement additional security measures on their home computers. However, they identified that response efficacy and self-efficacy were moderate predictors of individuals' intention to implement additional security measures. \citeauthor{klobas2019perceived}~\cite{klobas2019perceived} study on security risk's influence on smart home adoption (survey, $n=405$) observed the perceived risk as a determinant of smart home adoption intentions. \citeauthor{white2017analysis}~\cite{white2017analysis} used a survey with 945 adult participants to investigate different factors that affect computer security protective behaviors and perceived security incidents. \citeauthor{dupuis2018help}~\cite{dupuis2018help} used a mixed-method study under the framework of PMT, using analysis of data from 500 customer reviews of ten IoT devices, 18 interviews, and a large-scale AMT-based online survey with 1,006 valid response, to study the lack of privacy-risk awareness. They found that home users would engage in different privacy protected mechanisms if they are simple to understand and cheap to use. \citeauthor{George2021Theory}~\cite{George2021Theory} reiterated this fact in their survey ($n=219$), when they found that low awareness of risk coupled with self-efficacy hinders the users from addressing the existing privacy risks. \citeauthor{Anderson2010practicing}~\cite{Anderson2010practicing} used PMT to examine security behaviors of 101 participants in a survey with 594 home computer users and an experiment with 101 participants. They concluded that users' security behaviors were influenced by an individualized message focusing on the benefits of good security behaviors.

Two Papers extended the functionalities of PMT for their investigation. \citeauthor{Tsai_2016_Understanding_Conceptual}~\cite{Tsai_2016_Understanding_Conceptual} examined how PMT factors predict users' security intentions (survey, $n=988$). They extended the original PMT theory by including commonly neglected variables such as threat susceptibility, prior experience with a safety hazard, coping self-efficacy, to understand threat perspectives of home computer users while being online. They found several factors such as gender, age, threat severity, prior experience, coping self-efficacy, personal responsibilities amongst others, that are significantly co-related with users' security intentions. \citeauthor{thompson2017security}~\cite{thompson2017security} (survey, $n=629$), included the social and peer influence, psychological ownership and metrics on actual behaviors to measure the effectiveness of these factors on user behaviors under different contexts. Their results demonstrated that users behaved differently under both contexts (personal computers and mobile devices). Their findings echoed the PMT theory by proving the fact that perceived vulnerability, self-efficacy and response cost all played an important role in determining users' security behaviors.

The \textbf{Theory of Planned Behavior (TPB)} proposed by \citeauthor{ajzen2011theory} in \citeyear{ajzen2011theory}~\cite{ajzen2011theory} explains that individuals depend on intention to behave in a certain way and their ability to control that intention. \citeauthor{yang2017user}~\cite{yang2017user}added six external variables to TPB to build a comprehensive new model and validated the model with data collected from 216 survey participants. The results echoed the core concept of TPB that attitude, subjective norm and PBC (perceived behavioral control) are positively related to behavioral intention of the user. \citeauthor{guhr2020privacy}~\cite{guhr2020privacy} developed a research model based on several theoretical models, including TPB, to measure the effect of privacy concerns on the smart device usage by home users. The study (survey, n=$256$) applied the partial least squared structural equation modelling (PLS-SEM) to identify four essential elements to represent privacy concerns, including secondary use of personal information, perceived surveillance, perceived intrusion, and awareness of privacy practice. \citeauthor{Ferraris_2020_TrustModel_HomeNetwork}~\cite{Ferraris_2020_TrustModel_HomeNetwork} as suggested in \ref{subsubsec:RQ1}, put-forward a holistic trust model to improve security at home.

Some other old and new conceptual frameworks include the Technology Threat Avoidance Theory (TTAT) for testing the IT Threat avoidance~\cite{liang2010understanding}, negative-perception modelling for identifying barriers to smart home usage by the elderly user~\cite{pal2019embracing}, the privacy calculus theory to measure the relationship between perceived privacy risk and the willingness to share privacy information \cite{KIM_2019_Willingness}, Innovation Resistant theory (IRT) and Multidimensional Development Theory (MDT) to examine privacy concerns by \citeauthor{Pal2021Theory}~\cite{Pal2021Theory} and the new model Perceived surveillance of conversation (PSoC) developed by \citeauthor{Frick2021Theory}~\cite{Frick2021Theory} to determine the cause of privacy concern. Some of the new frameworks include the Stimuli-organisms-responses (S-O-R) framework for measuring the balancing role of negative emotions such as anger, anxiety between privacy concerns and behaviors~\cite{Park_2020_UserCognitive_Conceptual}, vulnerability-privacy concern-resistance (VPR) framework for explaining how users' resistance to the adoption of new technology is affected by their privacy concerns and their perception of their vulnerabilities \citeauthor{lee2020home}~\cite{lee2020home}. 

Although old and new theoretical frameworks have been used/developed to explore home users' security and privacy attitudes, practices and concerns, studies on using such \textbf{frameworks to analyze security and privacy aspects on the use of smart devices such as smart speakers} are largely missing from the papers we covered. In addition, it seems that such frameworks have not been incorporated into relevant ontologies to support related research in a more holistic manner.

\section{Discussions \& Recommended Research Directions}
\label{sec:discussions}

The results of our meta-review and SLR showed active and extensive research on different topics in the broad area of user perspectives of security and privacy aspects of home networks. However, our work also revealed many research gaps, indicating that more research is still required. In this section, we summarize our core findings around seven recommended future research directions. Note that some can be mapped to a single RQ of our SLR, but others cross-cut several RQs.

\subsection{Co-existence of multiple connected devices in a single home}

As mentioned in the Introduction Section, the average number of connected devices in an average household in most Western countries is over seven~\cite{Statista_2022_home_devices_household_worldwide2020}. The existence of many households with multiple home devices calls for more research on the role of co-existence of multiple devices in a single home. Such needs have been met by some research~\cite{boussard2018future, sikder2019multi}, but with limited depth and breadth. We observed three main research gaps. First, most research that has been conducted focused on standalone devices or a specific type of devices such as smart speakers~\cite{chalhoub2020alexa,abdi2019more, chhetri2019eliciting, ParkLim_2020_UserExpectations}, smart phone~\cite{cannizzaro2020trust}, and activity sensors~\cite{reeder2020older}. Although recent research has explored multiple devices in a connected home, especially data flows in a connected home~\cite{bugeja2019design, bugeja_2020_is, kilic2022cardboard}, more research is still needed to investigate \textbf{how different types of home devices interact with each other}, e.g., a smart doorbell with a smart alarm, how such interactions are perceived by the users and how they affect security and privacy of the home network as a whole. Second, the increasing number of home devices in a single home will \textbf{unavoidably complicate configuration and management of such devices}, including their security and privacy settings. Third, a large number of studies have concentrated on specific types of home devices, leaving \textbf{some types of home devices under-studied, especially different types of smart appliance}. However, the use of smart appliance at homes has been steadily increasing~\cite{Statista_2021a_smartAppliances, Statista_2021b_smartAppliances}, so more research on such devices is much needed.

\subsection{Multiple users in a single home}

According to the PRB (Population Reference Bureau)~\cite{PRB:household_size}, the average household size worldwide in 2020 was 4.0, suggesting that most home networks have multiple users. Considering frequent and occasional visitors (e.g., neighbors, relatives, friends, carers and nannies) to a household, the number of users can be even larger. There are also more complicated scenarios where the concepts of ``home'' and ``regular occupants'' are not clearly defined. For instance, some members of a household split their time between two or even more ``homes'' (e.g., university students and boarding school pupils live on campus during term time and go back to their parents' house during term breaks), some people living in the same neighborhood may share a broadband router where the ``home'' network covers multiple households (which we could call an ``extended home network''), and students sharing a multi-room house may see it as a ``pseudo-home''. Note that there can also be a hierarchical or graph-based structure among multiple home users, possibly device-dependent (e.g., a primary user of a home device is a secondary user of another home device). Research on the privacy and security issues in a multi-user home is steadily growing~\cite{zeng2019HomeUser, geeng2019Multi_User, huang_2020_amazon,bernd2020bystanders, yao2019privacy, Marky2021multi-user, Markey_2020_Idont_MultiUser,malkin2019privacy}. However, most studies in this area are focused on access control issues and power-play relationships between primary and secondary home users, overlooking the issues of an increasingly \textbf{hybrid and extended home occupants}~\cite{zhao2020unraveling}. Research on many aspects of multiple users in a single home, e.g., insider attacks and home users' perception, security and privacy aspects of an ``extended home network'' and a ``pseudo-home'' network, is missing from the current research literature.

\subsection{Multiple contexts and contextual factors/parameters}

Our SLR results showed evidence of research mainly in location-~\cite{mccreary2016context, Molina_2019_Contextual} or device-based contexts~\cite{tabassum2019perception, mcgill2017old}. Although home itself may be considered a specific context, home users actually use the home network and home devices in the home for many different purposes, leading to multiple different (sub-)contexts of home networking where security and privacy aspects have to be studied differently. For instance, when using traditional computing devices (personal computers) and ``smarter'' home devices, the context of use is very different. Similarly, when working from home, there is a mixture and overlap between the work and home contexts. Furthermore, when a user brings home devices (e.g., mobile devices and wearables) outside of the home for controlling home devices remotely, an ``extended home'' context is created. More generally, each unique home networking scenario and each specific type of home device could define a unique context, and the different subsets of home devices that work together for a specific purpose also define different contexts. In addition to contexts defined by \textbf{different usage scenarios and user intention/purposes}, some contexts are more overarching and should be considered part of other contexts, e.g., the legal context regarding data protection matters about home devices that collect personal data. Context-aware security and privacy is a significant research area and some past studies~\cite{rosa2015multi} have shown that concepts such as contextual histories dealing with the present and past contexts of the user can be used to enhance the competency and predict future contexts~\cite{da2016oracon} to adept user behaviors. However, as our SLR showed, research on different contexts and contextual factors/parameters is still relatively limited, and future work should venture into less-studied contexts.

\subsection{Data flows across multiple devices, multiple users and in multiple contexts}

Given the existence of multiple devices, multiple users and multiple contexts in a typical home network, and the complicated relationships between them, there can be complicated unidirectional and bidirectional data flows of different kinds, e.g., device-to-device, user-to-user, device-to-user, device-to-Internet, and user-to-Internet (the last two are about data flows between the home and the external world, mainly external online services and cloud servers on the Internet). These data flows might also differ in different contexts. Therefore, understanding such data flows is of particular importance for analyzing security and privacy issues and for developing more effective solutions. Researchers have been working around this topic~\cite{bugeja2018empirical, ren2019information, kilic2022cardboard} with a good deal of work focusing on device-to-Internet phenomena~\cite{tabassum2019perception, apthorpe2018spying}, but a more systematic endeavor is still lacking. Hence, more comprehensive research is needed to consider the complexities of data flows inside a home and from/to the external world. These might consider \textbf{different types of data flows between different entities in different contexts}, how they lead to security and/or privacy threats and risks, how home users perceive such data flows, how home users respond to any security or privacy concerns, and how technical or socio-technical solutions can be developed based on knowledge of such data flows. In addition, more experimental work is required to test ``hidden'' data flows that are not explicitly specified in user manuals of home devices and privacy policies of the manufacturers.

\subsection{Demographic factors}

Not surprisingly, our SLR revealed that gender and age are the two mostly studied demographic factors in the literature. Some researchers also looked at effects of other demographic factors such as ethnicity, knowledge, education level, income level, and disability, though with a limited depth. Furthermore, as mentioned in Section~\ref{subsubsec:RQ7}, 39\% of the studies were conducted in UK, USA and Western Europe, suggesting a need for \textbf{diversification of countries covered}. \citeauthor{chen2020Location}~\cite{chen2020Location} suggested that, ``geographical perspectives'' could be instrumental in deciding human behaviors and hence it is important that further research should consider less-studied areas. While some demographic factors have been less or not studied, even for more-studied factor such as gender, we have observed \textbf{conflicting results}, so more research is needed to consolidate our understanding.

\subsection{Home users' awareness, perceptions, attitudes, behaviors and practices}

Although a lot of work has been done on these aspects, the research gaps we identified in the previous subsections suggest that there are still several gaps to be filled by further research. Many factors, including household types, types of home networks including those with multiple geo-locations~\cite{zhao2020unraveling}, different user structures~\cite{zeng2019HomeUser, Markey_2020_Idont_MultiUser}, different usage contexts and scenarios, and different demographic factors, can affect home users' awareness, perceptions, attitudes, behaviors and practices. According to some past studies~\cite{gundu2013ignorance, binns2017Privacy, Knutzen2021Awareness, Bermejo2021Awareness}, behavioral practices could be influenced by the level of awareness. Although different awareness enhancement initiatives are steadily increasing to help users understand the nuances of security and privacy at home, they are still not effective for non-expert users. Therefore, more research is needed to look into better ways and methods of increasing privacy and security awareness of the users. Furthermore, a number of studies~\cite{kaaz2017understanding, topa2018usability} looked into issues of device configurations by users. These could be further investigated to address different issues, including weak authentication~\cite{tan2021secure} and multi-user authorization (automatic configuration)~\cite{bauer2020foresight}. Different ML algorithms including supervised classifiers~\cite{naeini2017privacy, barbosa2020privacy} and unsupervised learning algorithms such as LDA~\cite{barbosa2020privacy} have been used in past studies to both predict and analyze user behavior. However, further attention in this area is needed, especially in the context of \textbf{multi-user and multi-device home} to automatically learn and predict multi-user behavior and issues in a connected landscape. In addition to more studies on less-studied areas, some more \textbf{holistic approaches} are clearly needed, e.g., new taxonomies and ontologies that can cover different types of user perspectives and influencing factors.

\subsection{Theoretical and conceptual frameworks}

Our SLR has shown that past studies have considered the use and development of theoretical and conceptual frameworks, but mostly focusing on behavioral frameworks (e.g., PMT and TPB). On a more technical front, many taxonomies and ontologies have also been proposed~\cite{sarwar2018Taxonomy, Anwar2017SmartHome, solove2006taxonomy, heartfield2018taxonomy}, but they mostly have a limited scope and do not cover user perspectives sufficiently or not at all. More precisely, there is much less work developing theoretical and conceptual frameworks connecting home computing, IoT and smart home, user perspectives, security and privacy aspects, and other important factors. We argue that a more \textbf{advanced ontology needs developing to have a more holistic and comprehensive view of security and privacy of home networks}, covering a wide range of aspects including at least the following: traditional computing devices (including personal computers, routers and switches), smarter devices (including mobile devices, wearables, and IoT devices), physical and virtual network topologies, relevant software tools and online services (including firmware in hardware devices, mobile apps, smart speaker skills, online management tools and services, etc.), household and user structures, demographic factors, mappings between home devices to capabilities, different threats and defense mechanisms, different aspects of user perspectives (including awareness, attitude, perception, purposes and intention, behavior, activities, and practices). Developing such a comprehensive ontology is not trivial, and should be based on existing taxonomies and ontologies covering different areas.

\section{Conclusion}
\label{sec:Conclusion}

The purpose of this study is to conduct a systematic review of published papers on user perspectives of security and privacy aspects of a home network environment. It is evident from the results that this is quite a popular area of research and the number of studies, especially towards the later part of the last decade, has increased significantly in number and in depth. Despite many research papers published, the focus of most past studies is on issues and concerns arising from using a specific type of smart devices (i.e., smart speakers), their security and privacy practices and related decisions, and underlying factors such as user trust and access control issues. Few studies explored the issues of multiple types of connected devices inside a home network (including smart devices, traditional computing devices, multiple smart devices and gateway devices, etc.) or considered the fluid boundaries of a digital home. Additionally, some researchers also discussed multi-user related security and privacy concerns and behaviors. Research also highlighted the role of location- and device-specific contexts, and demographic factors, such as age and gender, in shaping users' security and privacy behaviors. Furthermore, the study collated theoretical and conceptual frameworks explaining the reasoning behind such users' behaviors and actions.   

Our work revealed a number of important research gaps and calls for more research in a range of key research areas, particularly around more holistic approaches (such as more advanced conceptual frameworks, especially a more comprehensive home networking and smart home ontology) considering multiple and inter-connected heterogeneous home devices, co-existence of several types of home users and other stakeholders, various contexts, data flows between different entities and in different contexts, and more demographic factors. In other words, we call for more future research to study the multi-dimensional complexity around security and privacy aspects of home networks and user perspectives, in order to make future home networks and smart homes more secure and privacy-friendly and meet people's needs better.

\bibliographystyle{ACM-Reference-Format}
\bibliography{main}

\end{document}